\documentclass[10pt,journal]{IEEEtran}
\usepackage{float}
\usepackage[table]{xcolor}
\usepackage{cite}      
\usepackage{graphicx}  
\usepackage{psfrag}    
\usepackage{hyperref}
\usepackage{cleveref}
\usepackage{placeins}
\usepackage{url}
\usepackage{stfloats}
\usepackage{times}
\usepackage{textcomp}
\usepackage{amsmath}   
\usepackage{amsfonts,amsthm,amssymb}
\usepackage{mathtools}
\usepackage{nicefrac}
\usepackage{multirow}
\usepackage[none]{hyphenat}
\usepackage{balance}
\usepackage{widetext}
\usepackage[caption=false,font=footnotesize]{subfig}
\usepackage{booktabs} % To thicken table lines

\usepackage{cite}
\usepackage[]{algorithmicx}
\usepackage{algpseudocode,algorithm}
\usepackage[switch]{lineno}

\usepackage{array}
\usepackage{fontenc}
\usepackage{color}
\usepackage{etoolbox}
\usepackage{bm}
\usepackage{comment}
\usepackage{algorithm}
\newcommand{\LRT}[2]{%
  \mathrel{\mathop\lessgtr\limits^{#1}_{#2}}%
}
\definecolor{LightCyan}{rgb}{0.88,1,1}

\begin{document}
%\linenumbers
\title{Doppler-Resilient 802.11ad-Based Ultra-Short Range Automotive Joint Radar-Communications System}

\author{Gaurav Duggal$^{\dagger}$, Shelly Vishwakarma$^{\dagger}$, Kumar~Vijay~Mishra and Shobha~Sundar~Ram
\thanks{$^{\dagger}$G. D. and S. V. are co-first authors.}% <-this % stops a space
\thanks{G. D., S. V. and S. S. R. are with the Indraprastha Institute of Information Technology Delhi, New Delhi 110020 India. E-mail: \{gaurav17091,shellyv,shobha\}@iiitd.ac.in.}%
\thanks{K. V. M. is with the IIHR - Hydroscience and Engineering, The University of Iowa, Iowa City, IA 52242 USA. E-mail: kumarvijay-mishra@uiowa.edu.}% <-this % stops a space
}

% make the title area
\maketitle
\begin{abstract}
% 75 word abstract for TAES
We present an ultra-short range IEEE 802.11ad-based automotive joint radar-communications (JRC) framework, wherein we improve the radar's Doppler resilience by incorporating Prouhet-Thue-Morse sequences in the preamble. The proposed processing reveals detailed micro-features of common automotive objects verified through extended scattering center models of animated pedestrian, bicycle, and car targets. Numerical experiments demonstrate $2.5$\% reduction in the probability-of-false-alarm at low signal-to-noise-ratios and improvement in the peak-to-sidelobe level dynamic range up to Doppler velocities of $\pm144$ km/hr over conventional 802.11ad JRC.
\end{abstract}

\section{Introduction}
\label{Sec:Intro}
During past few years, autonomous vehicles or self-driving cars have witnessed enormous development in vehicular control \cite{bengler2014three}, environmental sensing \cite{slavik2019phenomenological}, in-vehicle entertainment \cite{kong2017millimeter}, efficient resource utilization \cite{mishra2017auto}, and inter-vehicular synchronization \cite{gerla2014internet}. An ongoing challenge is target detection and recognition in order to avoid road accidents and boost automotive safety. Conventional target detection techniques use sensors such as lidar, camera, and infrared/thermal detectors. However, only radar offers the advantage of robust detection in adverse vision and weather conditions \cite{slavik2019phenomenological}. Currently, millimeter-wave (mm-Wave) ($77$ GHz) automotive radars are the preferred radar technology for target detection because they have wide bandwidths (\texttildelow$4$-$7$ GHz) and, hence, high range resolution \cite{mishra2019toward,mishra2017performance,mishra2018cognitive}. 

A concurrent development in intelligent transportation systems is the evolution of various vehicle-to-X (V2X) communication frameworks including vehicle-to-vehicle (V2V), vehicle-to-infrastructure (V2I), and vehicle-to-pedestrian (V2P) paradigms \cite{harding2014vehicle}. The overarching objective of these frameworks is to encourage sharing of road and vehicle information for applications such as environmental sensing, collision avoidance, and pedestrian detection. In the mm-Wave band, the IEEE 802.11ad protocol at unlicensed $60$ GHz has been identified as a potential candidate for these communications because of high throughput advantages arising from wide bandwidth \cite{mishra2019toward}. %The IEEE 802.11ad is also a crucial feature of future fifth generation (5G) networks whose high speed and data processing capabilities make it highly desirable for connecting autonomous vehicles \cite{kong2017millimeter}.

More recently, there is active research thrust towards combining automotive radar and communication functionalities on a single carrier 802.11ad wireless framework; the primary benefits being sharing of the common spectrum and hardware resources by the two systems (as already demonstrated at other bands \cite{mishra2019toward,dokhanchi2019mmwave,ayyar2019robust}). The \textit{stand-alone} conventional FMCW and noise waveform currently employed in automotive radars are not optimized for joint radar-communications (JRC). When \textit{modified} for use as a joint waveform, e.g. as shown in some of our previous works \cite{dokhanchi2019mmwave,alaee2019discrete,slavik2019cognitive}, FMCW and noise waveforms lead to largely an enhancement in a \textit{radar-centric} performance (e.g. probability of detection); the communications performance remains sub-optimal. Further, popular existing wi-fi protocols, especially at millimeter-wave, do not employ these signals. In this context, a \textit{communications-centric} JRC architecture based on 802.11ad not only exploits only a standardized mm-Wave communications protocol but has been shown in recent works \cite{kumari2015investigating,muns2017beam,mishra2019toward} to be suitable for JRC automotive system.

The 802.11ad-based V2V JRC was first proposed in  \cite{kumari2015investigating,grossi2018opportunistic}. The corresponding V2I application has been explored recently in \cite{muns2017beam} for radar-aided beam alignment to improve mm-Wave V2I communications. These works exploit the 802.11ad link to estimate ranges and Doppler velocities of automotive targets that are modeled as simple \textit{point} scatterers. This representation based on Swerling-0 model \cite{richards2005fundamentals} is appropriate for medium and long-range automotive radar applications where the far-field condition between the sensor and the target is sufficiently satisfied. In practice, however, 802.11ad is unsuitable for longer ranges because significant signal attenuation at $60$ GHz arising from oxygen absorption severely restricts the maximum detectable radar range \cite{richards2005fundamentals}. Therefore, it is more useful to employ 802.11ad-based JRC for \textit{ultra-short} range radars (USRRs). These sensors operate below $40$ m range \cite{jansen2017automotive,narnakaje2019automated} and have garnered much interest for applications such as blind spot warning, closing vehicle detection, lane change, park distance control, parking lot measurement and automatic parking assistance \cite{jansen2017automotive}.

\begin{figure}[t]
\centering
\includegraphics[width=0.75\columnwidth]{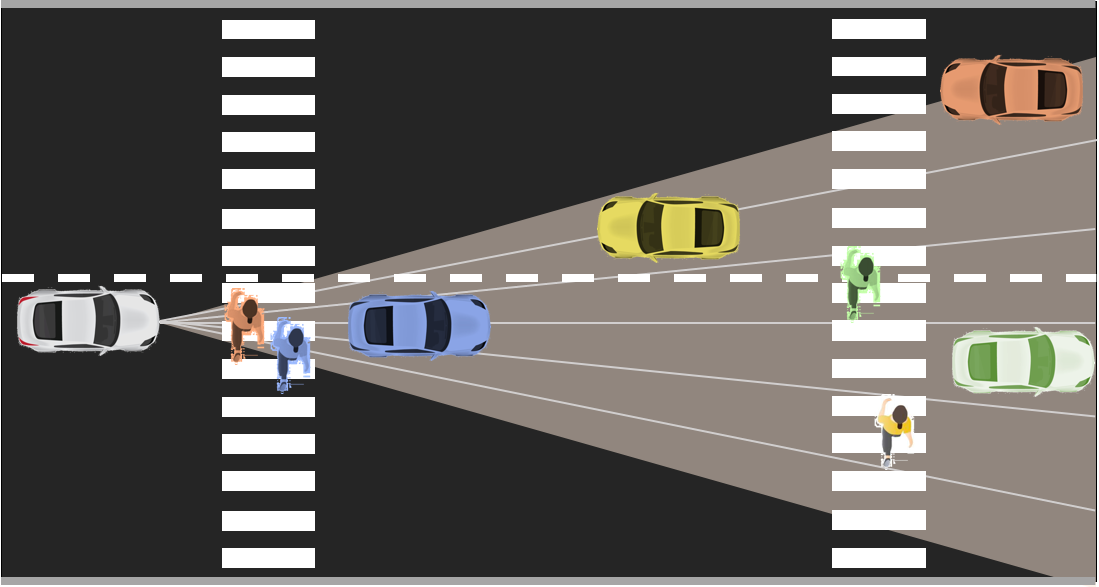}
\caption{Simplified illustration of an automotive radar scenario. The white car on left is mounted with the radar whose approximate coverage is indicated by the gray triangular area. The solid white lines within this area indicate azimuth bins. The targets at close range (blue and orange pedestrians; blue and yellow cars) occupy several cross-azimuth bins. Such targets are modeled as multiple point scatterers, each of which exhibits micro-motion. On the contrary, the targets at long range (green and yellow pedestrians; green and orange cars) fill up only a part of a single azimuthal bin and their micro-motions are indistinguishable.}
\label{fig:usrr}
\end{figure}

Employing 802.11ad for USRRs leads to a second problem. When the target is located within a close range of a high-resolution radar, the received signal is composed of multiple reflections from different parts of the same object \cite{dokhanchi2019mmwave} (Fig.~\ref{fig:usrr}). When the automotive target moves, these point scatterers may often exhibit \emph{micro-motions} besides the gross translational motion of the dynamic body. Examples include the swinging motion of the human arms and legs and the rotation of the wheels of a car. These micro-motions give rise to \emph{micro-Doppler} (m-D) features captured in joint time-frequency transforms \cite{chen2000analysis, chen2006micro} and/or \emph{micro-range} (m-R) features  captured in high range-resolution profiles (HRRPs) \cite{ram2009simulation}. These signatures are usually both distinctive and informative and have been used for target classification especially in the case of pedestrians \cite{bjorklund2011millimeter}. This \textit{extended} automotive target model is more general. But previous works on 802.11ad JRC represent targets as only point scatterers because, as we explain next, conventional 802.11ad waveform performs poorly in detecting both bulk and micro-motions.

The physical layer of IEEE 802.11ad protocol transmits control (CPHY), single carrier (SC) and orthogonal frequency-division multiplexing (OFDM) modulation frames at chip rates of $1.76$ GHz and $2.64$ GHz, respectively. Every single CPHY and SCPHY frame is embedded with a 2172-bit short training field (STF), a 1152-bit channel estimation field (CEF), 64-bit header, data block and a beamforming training field (omitted, for simplicity, in Fig.~\ref{fig:802.11.ad frame}). The CEF consists of two 512-point sequences $Gu_{512}[n]$ and $Gv_{512}[n]$ which encapsulate \textit{Golay complementary pairs} \cite{mishra2019toward}. These paired sequences have the property of perfect \emph{aperiodic} autocorrelation which is beneficial for communication channel estimation \cite{mishra2017sub} and radar remote sensing \cite{kumari2015investigating}. 
%----------------------------------------
%\begin{figure}[t]
%\centering
%\includegraphics[scale=0.4]{ADAS.eps}
%\includegraphics[width=3.0in,height = 1.2in]{SCPHY.eps}
%\caption{Short range (40m) automotive radar applications for advanced driver assistance systems}
%\label{fig:ADAS}
%\end{figure}
%-------------------------------------

The 802.11ad-based radars proposed in \cite{kumari2015investigating,muns2017beam,grossi2018opportunistic} harness the zero sidelobe attribute of 802.11ad Golay pairs during the matched filtering stage of the radar receiver to estimate the target's location in a delay-Doppler plane \cite{mishra2012signal,mishra2012frequency}. However, the perfect auto-correlation property of Golay pairs holds strictly for only static targets. When the target is moving, the Doppler phase shift in the received signal causes a deterioration in the pulse compression output leading to large non-zero side lobes \cite{pezeshki2008doppler}. This effect is accentuated for multiple moving point targets as well as a single extended target with multiple point scatterers moving at different velocities. 
A large body of literature exists on designing single polyphase sequences \cite{levanon2004radar} as well as generalizations of complementary Golay waveforms \cite{sivaswamy1982self} to exhibit Doppler tolerance \cite{levanon2017complementary}. In particular, \cite{pezeshki2008doppler} employed Prouhet-Thue-Morse sequence \cite{allouche1999ubiquitous} to design Doppler-resilient Golay complementary pairs which are free of range sidelobes at modest Doppler shifts. Such a sequence is appropriate for detection of micro-motion signatures. In this work, we utilize the $Gu_{512}$ field to construct Doppler-resilient Golay complementary sequences across multiple 802.11ad packets and show that their performance in detecting the m-D and m-R signatures of extended automotive scatterers supersedes that of the standard 802.11ad waveform. 

We presented preliminary results with simplistic target models with non-fluctuating radar cross-section (RCS) along constant velocity straight line trajectories in \cite{duggal2019micro}. In this work, our main contributions are:\\
\textbf{1) Extended target investigation for 802.11ad JRC.} We present realistic simulation models of automotive targets  accounting for size, shape, material and aspect properties along complex trajectories involving acceleration from start, driving turns and returning to halt. Specifically, we consider the following common targets: a small car, a bicycle and a pedestrian. We realistically animate all three targets independently along a complex trajectory within the maximum unambiguous range of the radar. The pedestrian is animated through motion capture (MoCap) data while the dynamics of the car and bicycle are realistically modeled using the PyBullet animation software \cite{pyBullet}. Then we integrate the animation model of the target with an electromagnetic scattering center model to obtain the time-varying radar returns. Specifically, we model the car as a cluster of triangular plates with point scatterers on its body and four wheels; the bicycle is modeled with cylinders and the pedestrian is represented with ellipsoidal body parts and corresponding point scatterers. \\
\textbf{2) 802.11ad evaluation with realistic m-D and m-R automotive models at ultra short-ranges.} We retrieve the HRRPs and the Doppler spectrograms from the extended targets using \emph{standard} and the \emph{modified} Doppler-resilient Golay (SG and MG, respectively) waveforms embedded within the IEEE 802.11ad packets. The non-rigid dynamics of each of these targets - motion of the arms and legs of the pedestrian; motion of the wheels, handle bar, pedals in case of vehicles (car or a bicycle) - give rise to distinctive m-R and m-D features in the range-time and Doppler-time signatures. While m-D signatures have been extensively simulated for pedestrians \cite{ram2008simulation, ram2009simulation, ssram2010Sim}, this work - to the best of our knowledge - is the first to present simulated m-D and m-R signatures of vehicles, apart from also evaluating all of them for 802.11ad for the first time.\\
\textbf{3) Improvements over standard 802.11ad in Doppler tolerance and sidelobe suppression.} Our results with the modified Golay complementary sequence show an improvement of approximately 20 dB in the range side-lobe suppression over the standard protocol. Interestingly, these side-lobe levels are retained up to Doppler velocities of $\pm 144$ km/hr which is well beyond the maximum target speed for urban highways. \\
\textbf{4) Detection performance evaluation.} %for varying signal to noise ratios.}
We study the impact of the range sidelobe suppression on the radar operating curves for the standard and the modified Golay waveforms. When radar transmits the modified Doppler-resilient waveform, we observed a significant reduction in the probability of false alarm ($P_{\textrm{fa}}$) for low to moderate signal-to-noise ratios (SNRs) due to the lower sidelobe level along the range dimension. Under the condition of a constant false alarm rate, this results in an improved probability of detection ($P_\textrm{d}$) for the modified signal over the standard waveform. In fact, the new waveform is able to tolerate a minimum SNR of $-9.4$ dB in order to achieve a $P_\textrm{d}$ and $P_{\textrm{fa}}$ of $99\%$ and $10^{-6}$, respectively. In comparison, the standard sequence needs a much higher minimum SNR of $+2$ dB to achieve the same performance.

The paper is organized as follows. In the following section, we describe the signal model of 802.11ad-based radar and introduce our proposed Doppler-resilient link. In Section~\ref{Sec:MeasResults}, we present Doppler radar signatures of common automotive targets at short ranges as measured by an actual radar. In Section~\ref{Sec:TargetModels} we present the models of the three targets for the 802.11ad-based radar. We validate our methods through numerical experiments in Section~\ref{Sec:Experiments} and conclude in Section~\ref{sec:Conclusion}. 
\section{Signal Model}
\label{sec:Theory}
The range and Doppler estimation methods using the SCPHY CEF field of standard 802.11ad are described in \cite{mishra2017sub,muns2017beam,kumari2015investigating}. In \cite{grossi2018opportunistic}, estimation of target parameters using the CPHY frame has been mentioned. In the following, we introduce the radar signal model based on 802.11ad SCPHY that we have adapted for extended targets. %\\
%\textbf{Classical 802.11ad-based target localization}: 
\subsection{Classical 802.11ad-based target localization}
\label{subsec:std_tgt}
%----------------------------------------
\begin{figure}[t]
\centering
\includegraphics[width=0.60\columnwidth]{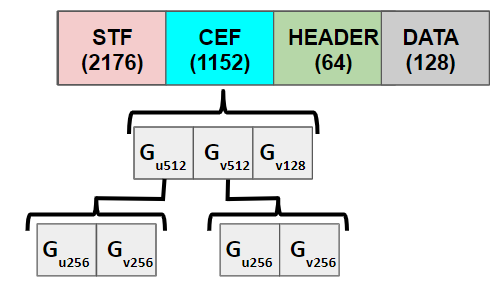}
\caption{Structure of the SCPHY IEEE 802.11ad frame which consists of the preamble (CEF and STF), a header, data blocks (BLK) and optional training fields (omitted). The CEF contains $G_{u512}$, $G_{v512}$ each of which comprises of a $256$ length Golay complementary pair. The numbers in parenthesis represent the length of the sequence.}
\label{fig:802.11.ad frame}
\end{figure}
%-------------------------------------
%----------------------------------------
\begin{figure}[t]
\centering
\includegraphics[width=1.0\columnwidth]{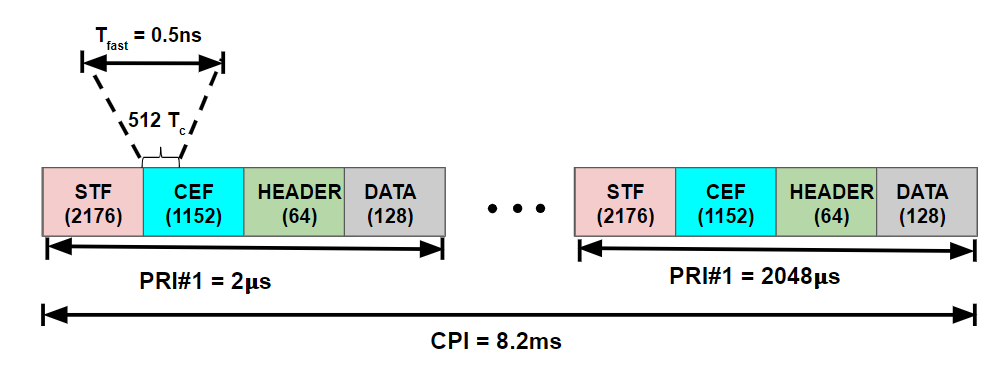}
\caption{Proposed Doppler-resilient Golay sequence in the CEF across multiple packets in a coherent processing interval. Consecutive packets contain one member of the Golay pair in the CEF field $Gu_{512}$ that is $512$-bit with time interval $T_c=0.5$ ns. A total of 4096 packets are transmitted.}
\label{fig:modified frame}
\end{figure}
%----------------------------------------
A \emph{Golay complementary pair} has two unimodular sequences $G_{1,N}$ and $G_{2,N}$ of the same length $N$ such that the sum of their autocorrelations has peak of $2N$ and side-lobe level of zero,\par\noindent\small
\begin{align}
\label{eq:golaytimeprop1}
G_{1,N}[n]*G_{1,N}[-n] + G_{2,N}[n]*G_{2,N}[-n] = 2N\delta[n].
\end{align}\normalsize
In previous studies \cite{kumari2015investigating,mishra2017sub,grossi2018opportunistic,muns2017beam,duggal2019micro}, the Golay complementary pair members $Ga_{256}$ and $Gb_{256}$ are drawn from the CEF of the same packet (Fig.~\ref{fig:802.11.ad frame}). When these pairs are correlated at the receiver, the pulse repetition intervals (PRIs) for both sequences in the pair differ by a delay equivalent to the transmission time of one 256-bit sequence. Such a non-uniform PRI has a bearing on Doppler estimation but was ignored in the previous studies that investigated only macro-Doppler features. In this work, to keep the PRI same among all members of the Golay pair, we propose that the complementary sequences are of length 512 and embedded in the $Gu_{512}$ of CEF alternately in two consecutive packets as shown in Fig.~\ref{fig:modified frame}. %Each 802.11ad packet consists of a 2172-bit long STF, 1152-bit long CEF, 64-bit Header, a data block and optional training fields (which, for simplicity, we have omitted in Figs.~\ref{fig:802.11.ad frame}-\ref{fig:modified frame}). 
For the $p$th packet, the transmit signal is the 512-bit Golay sequence in CEF,\par\noindent\small
\par\noindent\small
\begin{align}
\label{eq:txseq}
s_{T}[n] &= G_{p,512}[n], \phantom{1}n=0, 1, \cdots, 511,
\end{align}\normalsize
where $G_{p,512}$ and $G_{p+1,512}$ are Golay complementary pairs. The discrete-time sequence $s_T[n]$ is passed through a digital-to-analog-converter (DAC) the output of which can be represented as a weighted sum of Dirac impulses:
\par\noindent\small
\begin{align}
s_T(t) &= \sum\limits_{n=0}^{511}s_T[n]\delta(t-nT_c),
\end{align}\normalsize
where $F_c = 1.76$ GHz $= 1/T_c$. This signal is then amplified to impart energy $E_s$ per symbol to the transmit signal. The amplifier output is passed through a transmit shaping filter $h_T(t)$ to obtain \par\noindent\small
 \begin{align}
 x_T(t) = \sqrt{E_s}(s_T * h_T)(t) =  \sum\limits_{n=0}^{511} s_T[n]h_T(t-nT_c).
 \end{align}\normalsize
The 802.11ad protocol specifies a spectral mask for the transmit signal to limit inter-symbol interference (ISI) \cite[section 21.3.2]{ieee2012phy80211ad}. We assume that $h_T(t)$ includes a low-pass baseband filter with an equivalent amplitude characteristic of the spectral mask. A common shaping filter has a frequency response $H_T(f)$ of the root raised cosine (RRC) filter \cite{molisch2012wireless}. At the receiver, another RRC filter $h_R(t)$ is employed such that the net frequency response is equal to a raised cosine (RC) filter, $H(f) = H_T(f)H_R(f)$. The RC filter is a Nyquist filter with the following time-domain property to avoid ISI,\par\noindent\small
\begin{align}
\label{eq:rc1}
h[n]=h(nT_c)=\begin{dcases}
1,\;n=0\\
0,\;n\neq 0\\
\end{dcases}.
\end{align}\normalsize
We formulate this as\par\noindent\small
\begin{align}
\label{eq:rc2}
h(t)\sum\limits_{k=-\infty}^{+\infty}\delta(t-kT_c)=\delta(t).
\end{align}\normalsize
This property only holds for the RC, and not the RRC filter. The baseband signal is then upconverted for transmission: $x(t) = x_T(t)e^{j2\pi f_ct}$, where $f_c$ denotes the carrier frequency. The duration of this transmitted signal is $512T_c$ where $T_c$ is approximately 0.5ns and the number of fast time samples ($N$) is therefore 512. If we assume that the data block consists of 16 Bytes and that there are no optional training fields then each packet is of $T_p = 2$ $\mu$s duration which corresponds to the pulse repetition interval. 

Assume that the radar transmits $P$ packets constituting one coherent processing interval (CPI) towards a direct-path extended target of $B$ point scatterers. Each $b^{th}$ point scatterer is characterized by a time-varying complex reflectivity $a_b$ and is located at range $r_b = c\tau_b/2$ and Doppler $f_{D_b} = \frac{2v_b}{\lambda}$. Here $c = 3\times10^8$ m/s is the speed of light, $\tau_b$ is the time delay, $v_b$ is the associated radial velocity and $\lambda$ is the radar's wavelength. The coefficient $a_b$ subsumes common effects such as antenna directivity, processing gains and attenuations including path loss. Ignoring the multi-path components, the reflected received signal at the baseband, i.e., after down-conversion, over the duration of 1 CPI is\par\noindent\small
\begin{align}
\label{eq:RxModel}
x_{R}(t) &= \sum\limits_{p=0}^{P-1}\sum\limits_{b=1}^{B} a_b(t) x_T(t-\tau_b-pT_p)e^{-\mathrm{j}2\pi f_{D_b} t} + z(t)\nonumber\\
&\approx \sum\limits_{p=0}^{P-1}\sum\limits_{b=1}^{B} a_b(t) x_T(t-\tau_b-pT_p)e^{-\mathrm{j}2\pi f_{D_b}pT_p} + z(t),
\end{align}\normalsize
where $z(t)$ is additive circular-symmetric white Gaussian noise. The last approximation follows from the fact that $f_{D_b} \ll 1/T_p$ so that the phase rotation within one CPI (\emph{slow time}) can be approximated as a constant. Sampling the signal at $F_c = 1/T_c$ yields $x_R[n] = x_R(nT_c)$,
\par\noindent\small
\begin{align}
\label{eq:RxSig}
x_R[n]  
&= \sum\limits_{p=0}^{P-1}\sum\limits_{b=1}^{B} a_b[n] x_T(nT_c-\tau_b-pT_p)e^{-\mathrm{j}2\pi f_{D_b}pT_p} + z(nT_c)\nonumber\\
&= \sum\limits_{p=0}^{P-1}\sum\limits_{b=1}^{B} a_b[n] s_T(nT_c-\tau_b-pT_p)e^{-\mathrm{j}2\pi f_{D_{b}}pT_p} + z[n],
\end{align}\normalsize
where we used Nyquist filter properties (\ref{eq:rc1})-(\ref{eq:rc2}).% in the last equality. 

%As mentioned earlier, the transmit signal embeds inidvidual Golay pair sequences in consecutive packets. In other words, $G_{1,512}$ and $G_{2,512}$ are Golay pairs and $G_{3,512}$ and $G_{4,512}$ are again Golay pairs and so on.
When the sampled signal from two consecutive packets is passed through matched filters of each Golay sequence, we exploit \eqref{eq:golaytimeprop1} %the perfect autocorrelation property 
to estimate the radar channel. For instance, correlation for the $p^{th}$ pair produces\par\noindent\small
\begin{align}
\hat{h}_{p}[n] &= x_R[n]*G_{p,512}[-n]\nonumber\\
\hat{h}_{p+1}[n] &= x_R[n]*G_{p+1,512}[-n]
\end{align}\normalsize
These outputs are added to return the channel estimate\par\noindent\small
\begin{align}
\label{eq:perfectCh}
\hat{h}[n] &= \frac{1}{1024}(\hat{h}_{p}[n] + \hat{h}_{p+1}[n])\nonumber\\
			   &\approx
\frac{1}{1024}\sum\limits_{p=0}^{P-1} \sum\limits_{b=1}^{B} a_b[n] \delta(nT_c-\tau_b-pT_p)e^{-\mathrm{j}2\pi f_{D_b}pT_p}\nonumber\\
               &+ z[n]*(G_{p,512}[-n] + G_{p+1,512}[-n]),
\end{align}\normalsize
where the last approximation is due to the assumption that the Doppler shifts are nearly identical for the two Golay sequences $G_{p,512}$ and $G_{p+1,512}$ to utilize zero side-lobe property of (\ref{eq:golaytimeprop1}). 

The range space is discretized into $N=512$ bins of $cT_c/2$ resolution (range bins, $r_n=\frac{cT_cn}{2}, n=0,\cdots,N-1$). Therefore, the HRRP, $\chi^{RT}_m[r_n]$, corresponding to each $m^{th}$ CPI, is obtained by computing the radar channel estimate ($\hat{h}_{p,m}$) through correlation for all $P/2$ pairs within the CPI,\par\noindent\small
\begin{align}
\label{eq:RTplot}
    \chi^{RT}_m[r_n]=\frac{1}{512P}\sum_{p=0}^{P-1}\hat{h}_{p,m}[n]
\end{align}\normalsize
We locate the $B$ peaks of the extended target on the two-dimensional delay-Doppler, $\chi^{RD}_m[r_n,f_D]$, corresponding to the $m^{th}$ CPI, by taking a $P$-point Discrete Fourier Transform (DFT) of the radar channel estimates for each Doppler shift bin, $f_D$,\par\noindent\small
\begin{align}
\label{eq:RDAmbig}
    \chi^{RD}_m[r_n,f_D] = \frac{1}{512P}\sum\limits_{p=0}^{P-1} \hat{h}_{p,m}[n]e^{\mathrm{j}2\pi f_D pT_p}.
\end{align}\normalsize
For each $m^{th}$ CPI, the peaks along the range axes for each Doppler bin are coherently summed to obtain the Doppler (or velocity) - time spectrogram\par\noindent\small
\begin{align}
\label{eq:DTplot}
    \chi^{DT}_m[f_D] = \sum_{n=0}^{511}\chi^{RD}_m[r_n,f_D].
\end{align}\normalsize
\subsection{Doppler-resilient 802.11ad}
%\textbf{Doppler-resilient 802.11ad}:
%-----------------------------------------------------
\begin{figure}[t]
\centering
\includegraphics[width=0.75\columnwidth]{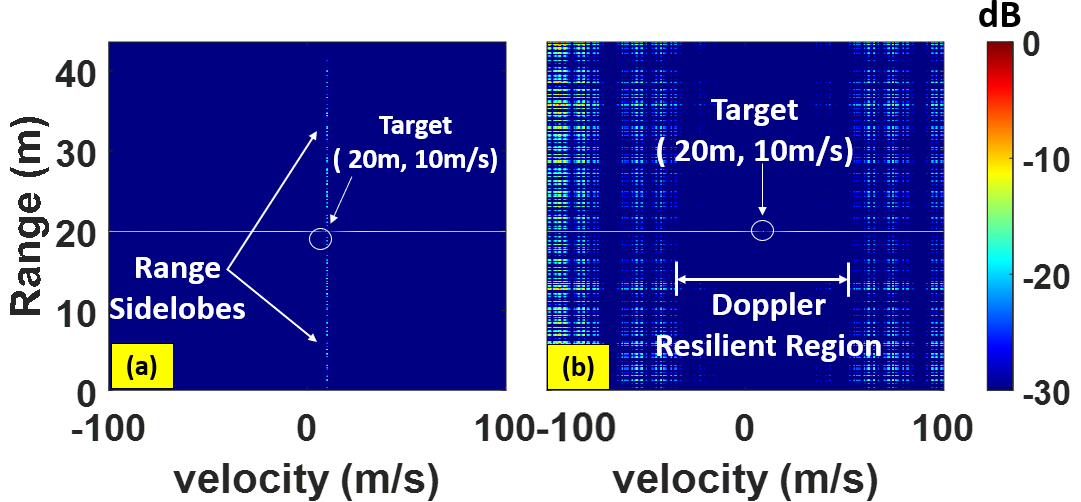}
\caption{Normalized ambiguity function for a point scatterer at range $20$ m moving with a constant Doppler of $10$ m/s for (a) standard Golay (SG) (b) Doppler-resilient Golay waveform.}
\label{fig:G1G2 ambiguity}
\end{figure}
%------------------------------------------------------
When a target is moving, the Doppler based phase shifts across the two PRIs may differ. For example, a point scatterer moving with a Doppler shift of $f_{D_b}$ will give rise to a phase shift of $\theta \approx 2 \pi f_{D_b} T_p$ between $\hat{h}_{p}$ and $\hat{h}_{p+1}$. In this case, the perfect autocorrelation would no longer hold, i.e.,\par\noindent\small
\begin{align}
\label{eq:Disturbance}
    \left(G_{p,N}[n]*G_{p,N}[-n]\right) + \left(G_{p+1,N}[n]*G_{p+1,N}[-n]\right)e^{-\mathrm{j}\theta} \ne 2N\delta[n],
    %\phantom{1}n=0, 1, \cdots, 511,
\end{align} \normalsize
resulting in high side-lobes along the range. For example, consider a simple nonfluctuating point scatterer of unit reflectivity at ($r_b = 20$ m, $v_b = 10$ m/s) over 1 CPI of $P = 4096$ packets. For this set of waveform parameters, Fig.~\ref{fig:G1G2 ambiguity}a plots the \textit{range-Doppler ambiguity function} ($\chi^{RD}(r_n,f_D)$) obtained by correlating the waveform with its Doppler-shifted and delayed replicas. The ambiguity function (AF) completely characterizes the radar's ability to discriminate in both range and velocity of its transmit waveform. 
The complementary Golay (standard Golay) AF shows a very high peak-to-sidelobe level of $-15$ dB at non-zero Doppler frequencies. This would result in \emph{high false alarms} especially at low SNRs. 

The limitation described above can be overcome by using  Doppler-tolerant Golay sequences such as the one proposed in \cite{pezeshki2008doppler}. Without loss of generality, assume $P$ be even and generate the Prouhet-Thue-Morse (PTM) sequence \cite{allouche1999ubiquitous}, $\{q_p\}_{p=0}^{\frac{P}{2}-1}$ which takes values in the set $\{0,1\}$ by following Boolean recursion\par\noindent\small
\begin{align}
    q_p &= 
    \begin{dcases}
    0,& \text{if}\; p=0\\
    q_{\frac{p}{2}},&\text{if}\;(p\; \text{modulo}\; 2) = 0\\
    \overline{q_{\frac{p-1}{2}}},&\text{if}\;(p \; \text{modulo}\; 2) = 1,
    \end{dcases}
\end{align}\normalsize
where $\overline{q_p}$ denotes the binary complement of $q_p$. As an example, when $P=16$, the PTM sequence is $q_0 = \{0,1,1,0,1,0,0,1\}$. 

Based on the values of $q_p$, we transmit the following Golay pairs: if $q_1=0$, then the complementary pair $G_{1,N}[n]$ and $G_{2,N}[n]$ are transmitted separately in two consecutive packets; if $q_2=1$, then the consecutive transmission consists of the complementary pair with $-G_{2,N}[-n]$ and $G_{1,N}[-n]$; and so on. In this manner, we transmit a sequence of Doppler-resilient Golay sequences over $P$ packets. The goal is to obtain 
a pulse train of Golay pairs such that\par\noindent\small
\begin{align}
\label{eq:drs1}
    \sum\limits_{p=0}^{P-1}e^{\mathrm{j}n\theta}(G_{p,N}[n]*G_{p,N}[-n])
    \approx f(\theta)\delta[n],
\end{align}\normalsize
where the function $f(\theta)$ does not depend on the time-shift index $n$ for some reasonably large values of $\theta$. The Taylor series approximation of %the left-hand-side of 
(\ref{eq:drs1}) around zero Doppler is\par\noindent\small
\begin{flalign}
\label{eq:drs2}
    &\sum\limits_{p=0}^{P-1}e^{\mathrm{j}n\theta}(G_{p,N}[n]*G_{p,N}[-n])\approx0(G_{0,N}[n]*G_{0,N}[-n])\nonumber\\ &
    + 1(G_{1,N}[n]*G_{1,N}-n])  + 2(G_{2,N}[n]*G_{2,N}[-n])  \nonumber\\ & + \cdots + (P-1)(G_{P-1,N}[n]*G_{P-1,N}[-n]).
\end{flalign}\normalsize
Using PTM sequence, the above summation can be made to approach a delta function. The key is to transmit a Golay sequence that is also complementary with sequences in more than one packet. For instance, when $P=4$, the PTM sequence dictates sending following signals in consecutive packets: $G_{1,N}[n]$, $G_{2,N}[n]$, $-G_{2,N}[-n]$ and $G_{1,N}[-n]$ for an arbitrary Golay pair $\{G_{1,N}[n], G_{2,N}[n]\}$. In such a transmission, not only the first and last two sequences are Golay pairs but also the second and fourth signals. This implies\par\noindent\small
\begin{flalign}
\label{eq:drs3}
    &\sum\limits_{p=0}^{3}e^{\mathrm{j}n\theta}(G_{p,N}[n]*G_{p,N}[-n]) \nonumber\\ 
    &\approx (G_{1,N}[n]*G_{1,N}[-n])+ 2(G_{2,N}[n]*G_{2,N}[-n]) \nonumber\\ 
    &+ 3(G_{3,N}[n]*G_{3,N}[-n])\nonumber\\
    &= 1((G_{1,N}[n]*G_{1,N}[-n]) + (G_{3,N}[n]*G_{3,N}[-n]))\nonumber\\ 
    &+ 2((G_{2,N}[n]*G_{2,N}[-n]) + (G_{3,N}[n]*G_{3,N}[-n]))\nonumber\\ 
    &=(2N+2(2N))\delta[n] = 6N\delta[n].
\end{flalign}\normalsize

For these Doppler-resilient Golay sequences, the resultant AF plot is nearly free of range sidelobes especially at low Doppler velocities. For the same target and waveform parameters as in Fig.~\ref{fig:G1G2 ambiguity}a, the corresponding AF plot for Doppler-resilient Golay sequences is shown in Fig.~\ref{fig:G1G2 ambiguity}b. Here, the peak-to-sidelobe level for Doppler-resilient Golay sequences is -$42$ dB. Hence, an improvement of $27$ dB is obtained over standard Golay sequences. We also note that the Doppler tolerance holds for target velocities up to approximately $\pm 40$ m/s (= $\pm 144$ km/hr) which is above most of the velocities encountered in automotive scenarios. Hence, this waveform is suitable for V2P and USRR applications. From here on, we refer to the Doppler-resilient Golay sequences as modified Golay (MG) and the original sequences presented in (\ref{eq:golaytimeprop1}) as standard Golay (SG). 

The IEEE 802.11ad physical layer (PHY) transmits single carrier (SC) modulation frames with $1.76$ GHz bandwidth at a carrier frequency of $60$ GHz. The range resolution is $0.085$ m, determined by the chip rate of $1.76$ GHz and the maximum range is $44$ m corresponding to 512 fast-time samples. 
%The maximum velocity likely to be encountered on roads corresponds to $150$ km/hour or approximately $40$ m/s. For a carrier frequency of $60$ GHz, this results in a maximum Doppler of $16$ KHz. 
In order to detect velocity accuracy of approximately $0.3$ m/s, we require a Doppler resolution of $122$ Hz and a CPI of $8.2$ ms. This implies transmission of $P=4096$ packets to form a single CPI. The maximum unambiguous velocity $\nu_{max}$ is determined by the PRI $T_p$: $\nu_{\text{max}} = \lambda/T_p$. Table~\ref{Table:parameters} summarizes the parameters of the proposed waveform. Better Doppler resolutions are possible by increasing the packet length. 
\section{Measurement Data Collection}
\label{Sec:MeasResults}
\begin{figure}[t]
    \centering
    \includegraphics[width=0.5\columnwidth]{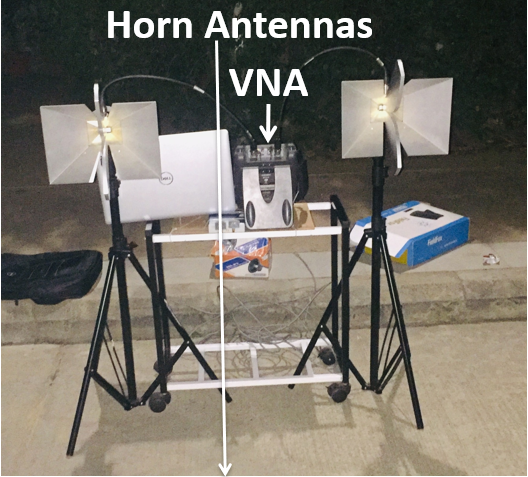}
    \caption{Monostatic radar configuration using vector network analyzer in narrowband mode and two horn antennas.}
    \label{fig:MeasSetUp}
\end{figure}

Unlike previous 802.11ad radar studies that assume only simple point targets at long ranges, real world automotive targets such as pedestrians, bicycles and cars appear as extended scatterers to the radar, more so at close ranges. We demonstrate this aspect with measured data in this section. We collected narrowband m-D data of a pedestrian, bicycle and a car using a monostatic radar consisting of a N9926A FieldFox vector network analyzer (VNA) and two horn antennas (HF907) (Fig.~\ref{fig:MeasSetUp}). The VNA was configured to carry out narrowband $S_{21}$ parameter measurements at a carrier frequency of $7.5$ GHz, with transmitted power set at $3$ dBm and sampling frequency at $370$ Hz (maximum frequency in the narrowband mode). The gain of both horn antennas is $10$ dBi. The return echoes of targets were recorded separately. 

%--------------------------------------------------
\begin{table}[t]
  \begin{center}
      \caption{Proposed radar parameters}
          \label{Table:parameters}
    \begin{tabular}{p{5cm}p{2.5cm}}
      %\hline \hline
      \toprule
      Parameter & Proposed V2P Radar  \\ 
      \midrule
      %\hline \hline
       Carrier frequency (GHz) & $60$ \\
       Bandwidth (GHz) & $1.76$ \\
       Range resolution (m) & $0.085$ \\
       Maximum unambiguous range (m) & $44$ \\
       Pulse repetition interval ($\mu$s) & $2$\\
       Velocity resolution (m/s) & $0.3$ \\
       Maximum unambiguous velocity (m/s) & $625$ \\
      \bottomrule
      %\hline \hline
    \end{tabular}
  \end{center}
\end{table}
%--------------------------------------------------
%--------------------------------------------------
%--------------------------------------------------

\begin{figure}[t]
\centering 
\includegraphics[width=0.8\columnwidth]{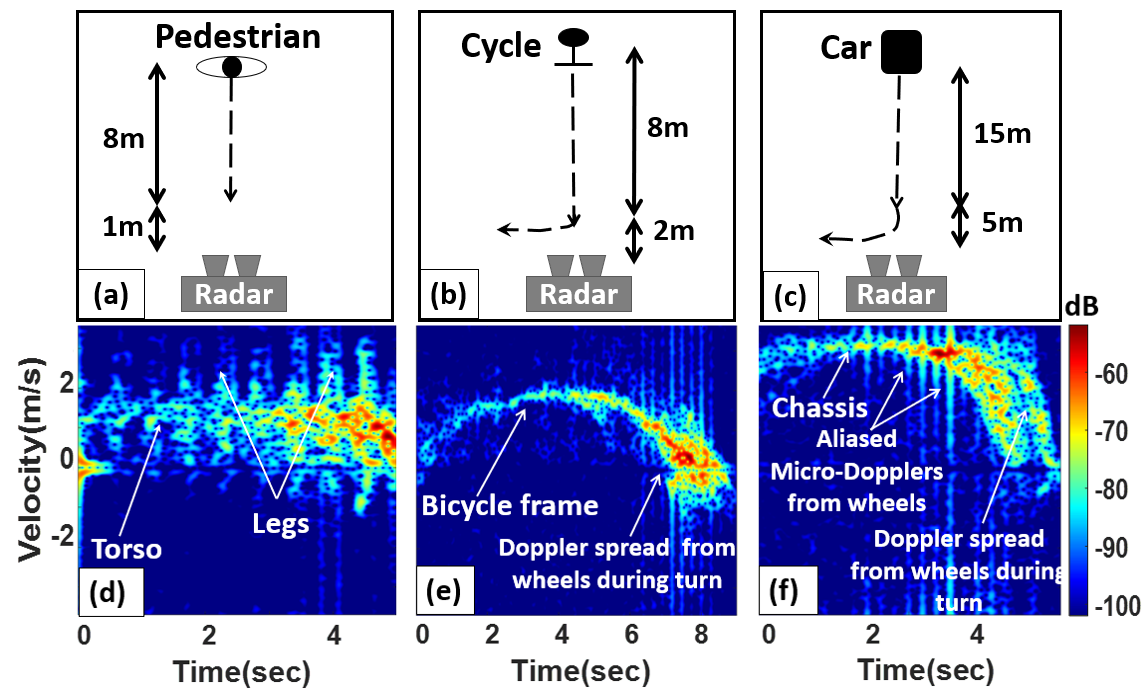}
\hfil
\caption{Trajectories followed by (a) pedestrian, (b) bicycle and (c) car during measurement data collection. The m-D spectrograms of (d) pedestrian, (e) bicycle and (f) car via STFT on $7.5$ GHz narrowband measurement data.}
\label{fig:MDspectrograms}
\end{figure}

The trajectories of the three targets are shown in Fig.~\ref{fig:MDspectrograms}. First, we consider a pedestrian of height $1.73$ m. The subject walks towards the radar from a distance of $8$ m (Fig.~\ref{fig:MDspectrograms}a) with an approximate speed of $1$ m/s. The resulting m-D spectrogram (Fig.~\ref{fig:MDspectrograms}d) demonstrates that the pedestrian must be regarded as an extended target because m-D features from the torso, arms and legs are clearly visible on the radar. All m-Ds are positive when the pedestrian is approaching the radar. The swinging motions of the legs give rise to the highest Dopplers, followed by the arms and the torso. Next, we consider a bicycle target of height $1.1$ m, length $1.8$ m and wheels of $0.45$ m radius. The bicycle starts from a distance of 10m and then turns left $2$ m before the radar (Fig.\ref{fig:MDspectrograms}b). The corresponding spectrogram (Fig.\ref{fig:MDspectrograms}e) indicates that when the bicycle is moving straight towards the radar, only the m-Ds from its frame are visible. However, when it executes a turn before the radar, multiple Doppler components arise from this single target. Besides the translational motion, % of the bicycle, 
the rotational motion of the two wheels turning at different velocities with respect to the radar; the motion of the pedals, and the small adjustments of the handle bars required for maintaining the balance of the bicycle also produce m-Ds. These features are best observed during $6$-$9$ s.  

%------------------------------------------------------
%------------------------------------------------------
Finally, we consider a small size car (Hyundai Grand I10) of dimensions $3.765$ m $\times$ $1.66$ m $\times$ $1.52$ m and wheels of radius $0.36$ m. The car moves from a distance of $20$ m from the radar and then turns left before the radar at a distance of about $5$ m (Fig.~\ref{fig:MDspectrograms}c). The car chassis %of the car 
moves with an average speed of $3$ m/s which generates translational Doppler. But the rotational motion of the wheels introduce m-Ds (Fig.~\ref{fig:MDspectrograms}f). Any point on the wheel circumference moves with a cycloidal motion.

\begin{figure}
    \centering
    \includegraphics[width=0.25\columnwidth]{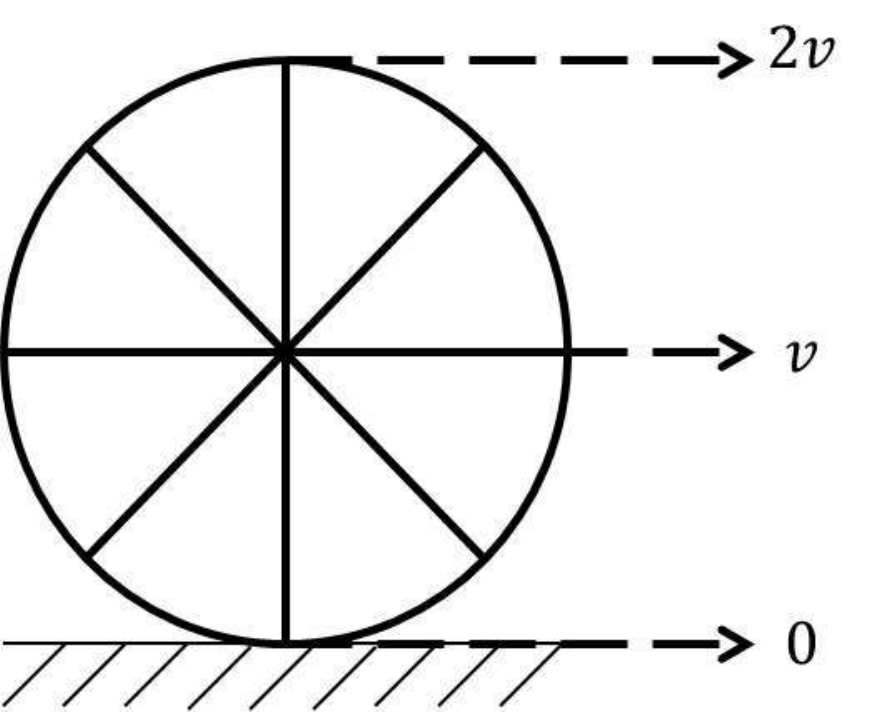}
    \caption{Velocity of point scatterers distributed along a rotating wheel}
    \label{fig:WheelVelocity}
\end{figure}

If the speed of the center of the wheel is ${\mathbf{v}}$ m/s, then the speed of the top of the wheel is $2{\mathbf{v}}$ m/s while the speed of the base of the wheel is set to zero Doppler because of friction (Fig.~\ref{fig:WheelVelocity}). Therefore, the m-Ds from the four wheels are spread from 0 to twice the mean Doppler from the chassis when the car moves in a straight line. Note that the angular velocities of the four wheels are usually identical in this scenario. However, depending on the path taken by the vehicle, the radial velocity components of the four wheels with respect to the radar may differ resulting in slightly varying m-D values. 
%------------------------------------------------------
%------------------------------------------------------

The returns from the wheels are usually much weaker than the strong Doppler from the chassis and, hence, are visible only when the car is near the radar (Fig.~\ref{fig:MDspectrograms}c). Due to the limited sampling frequency of the radar receiver ($370$ Hz), some aliased m-D components from the rotation of the wheels at the lower frequencies also show up. When the car is turning before the radar, then the right and left wheels turn at different radii resulting in very different radial velocities. This results in a large m-D spread that appears during $4$-$6$ s in the spectrogram. 

The measurement results show that typical automotive targets are extended targets at short ranges, resulting in distinctive m-D spectrograms. They, therefore, engender similar m-R features in HRRPs generated with broadband radar data. In the following section, we discuss the modeling of these extended target models of pedestrians, bicycles and cars before applying our new Doppler-resilient waveform and its associated processing. 
\section{Extended Target Models}
\label{Sec:TargetModels}
There are multiple methods for generating animation models of dynamic bodies \cite{ram2008simulation}. In this work, we derive the animation data of pedestrians from motion capture (MoCap) technology and use a physics based simulator to model the motion of a car and bicycle. Use of elementary shapes is well documented in radar simulations of humans (see e.g. \cite{ram2008simulation, ssram2010Sim}).
\subsection{Animation model of car and bicycle}
%\textbf{Animation model of car and bicycle}:
%------------------------------------------------------
\begin{figure}[t]
\centering 
\includegraphics[width=0.9\columnwidth]{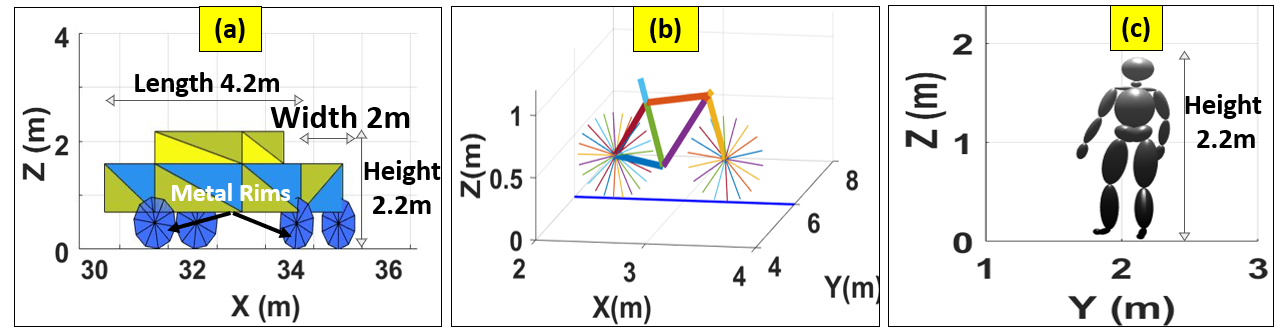}
\caption{Scattering center model of (a) a car, (b) a cycle and (c) a pedestrian.}
\label{fig:PointScattererModel}
\end{figure}
%------------------------------------------------------
We employed pyBullet - a Python based open source software development kit (SDK) - for generating motion data of a car and bicycle \cite{coumans2017}. PyBullet uses Bullet Physics, a physics-based animation package, for describing motions of dynamic bodies \cite{kokkevis1996user}. In this environment, each vehicle is designed as a collection of interconnected rigid bodies such that they do not undergo any type of physical deformation during motion. We modeled the car with a lateral wheel axle length of $2$ m, front axle to rear axle length of $3.5$ m and a wheel radius of $0.48$ m as shown in Fig.~\ref{fig:PointScattererModel}a. We simulated \emph{front wheel driving} of the car by considering the root of the compound body at the center of the axle connecting the two front wheels of the car. The root has six primary degrees of freedom (DOF) - translation along the three Cartesian axes and rotation around the same axes. A user \emph{drives} the vehicle at the desired speed and along the desired trajectory by prescribing specific kinematic trajectories to the root.  Secondary parts such as wheels are connected to the primary body through joints or hinges. The simulator then computes forces and torques that actuate the secondary DOFs of the joints to follow freely based on forward dynamics. The resulting motions of all the bodies comprising the vehicle are constrained by a control system in the software to realize realistic animation of the vehicle at a frame rate of $60$ Hz. In the case of the bicycle, we consider a bicycle frame with a cross bar frame and two wheels as shown in Fig.~\ref{fig:PointScattererModel}b. For the sake of simplicity, we do not model the human rider on the bicycle. Using PyBullet, we obtain a realistic animation model of the two wheels, the bicycle frame and the front handle bars. 
\subsection{Animation model of pedestrians}
%\textbf{Animation model of pedestrians}: 
The animation data of a walking human was obtained from MoCap technology at Sony Computer Entertainment America \cite{ssram2010Sim}. The data describes the time-varying three-dimensional positions of a collection of markers distributed over the body of a live actor at a frame rate of $60$ Hz over a duration of $5$ s. There are $24$ markers located at the head, torso (both front and back), upper arms, hands, knees and feet. %We assume that t
These markers correspond to the point scatterers on the body of the pedestrian as shown in Fig.~\ref{fig:PointScattererModel}c. 
\subsection{Electromagnetic modeling with varying aspect angles}
%\textbf{Electromagnetic modeling with varying aspect angles}: 
We integrate the animation data of the pedestrian, bicycle and car with electromagnetic models of radar scattering using the primitive modeling technique \cite{ram2008simulation} which has been extensively employed for modeling radar returns from dynamic human motions \cite{ssram2010Sim}. We first interpolate the animation data from the video frame rate to the radar sampling rate. Then, each of the dynamic bodies is modelled as an extended target made of multiple primitives with point scatterers distributed along its body. The car is assumed to be composed of 56 triangular plates. The wheel-rims and car body are modeled as metallic plates. The windows, front windscreen and rear windscreen screen are modelled as glass. The radar cross section of each $b^{th}$ plate at each $n^{th}$ discrete time instance is\par\noindent\small
\begin{equation}
\label{eq:RCSTriangularPlate}
    \sigma_b[n] = \frac{4\pi A_b^2 \cos^2 \theta_b[n]}{\lambda^2} \left(\frac{\sin\left(k d_b \sin \frac{\theta_b[n]}{2}\right)}{k d_b \sin \frac{\theta_b[n]}{2}}\right)^4,
\end{equation}\normalsize
where $A_b$ is the area of the triangle, $d_b$ is the dimension of the triangle along aspect angle and $k=\frac{2\pi}{\lambda}$ is the phase constant for $\lambda$ wavelength \cite{ruck1970radar}. The aspect angle $\theta_b[n]$ is defined as the angle between the incident ray from the radar and the normal to the triangular plate.
\par In the case of the bicycle, we have modelled the Argon 18 Gallium bicycle. The bicycle frame consists of 9 metal cylinders of $6$ cm radius of differing lengths. The front wheel has 18 spokes and the rear wheel has 25 spokes, each of which are of $2$ mm radius and $34.5$ cm length.
The RCS of a cylinder \cite{ruck1970radar} of length $L_b$ and radius $a_b$ is:
\par\noindent\small
\begin{equation}
\sigma_b[n] = \frac{2\pi a_b L_b^2}{\lambda}\cos^2(\theta_b[n]) \left( \frac{\sin \left(k L_b \sin \theta_b[n]\right)}{k L_b \sin \theta_b[n]}\right)^2,
\end{equation}\normalsize
For the pedestrian, 21 different body parts - torso, arms and legs - are modelled as ellipsoids while the head is approximated as a sphere. The longest dimension of each ellipsoid spans the length of the bone in the skeleton structure of the MoCap data. These assumptions follow the well-established kinematic modeling of walking human originally developed in \cite{boulic1990global} which is capable of intuitively showing the details of human movements and ensuring the correct placement of the radar system. The radar cross section of an ellipsoid \cite{ruck1970radar} of length $H_b$ and radius $R_b$ is
\par\noindent\small
\begin{equation}
\sigma_b[n] = \frac{\frac{1}{4}R_b^4 H_b^2}{R_b^2\sin^2 \theta_b[n]+\frac{1}{4}H_b^2\cos^2 \theta_b[n]}.
\end{equation}\normalsize
The RCS fluctuates with time due to the variation in angle $\theta_b[n]$ between incident wave and the major length axis of ellipsoid. %We assume that 
The ellipsoids are made of single layer dielectric with a dielectric constant of 80 and conductivity 2.

Assuming the transmit power $P_t$ and antenna processing gains of the transmitter $G_{t}$ and receiver $G_{r}$ antennas are unity, then the strength of the scattered signal from the $b^{th}$ part of the extended target depends on the material properties, aspect angle and the position of the scattering center on the primitive with respect to the radar. We incorporate the material properties of the target into the RCS estimation through Fresnel reflection coefficient, $\Gamma$, for planar interfaces at normal incidence. The attenuation of $60$ GHz wave through the air medium is modeled through $\alpha$. 
All scattering centers may not be visible to the radar at each time instant because of shadowing by other parts of the same target or other channel conditions. Therefore, we incorporate stochasticity in the scattering center model by including a Bernoulli random variable of mean $0.5$, i.e. $\zeta_b[n] \sim \textrm{Bernoulli}(0.5)$, in the RCS. A point scatterer is \emph{seen} by the radar with a probability of 50\% at every time instant. The reflectivity of a primitive at any time sample is
\par\noindent\small
\begin{equation}
    \label{eq:RCSellipsoid}
    a_b[n] =\zeta_b[n] \Gamma(\theta_b[n]) \sqrt{\sigma_b[n]} \frac{e^{-2\alpha r_b[n]}}{r_b^2[n]}.
\end{equation}\normalsize
The numerical delay of each point scatterer is estimated from the range as $\tau_b[n]=\frac{2r_b[n]}{cT_c}$. The Doppler shift, arising from the change in position with respect to time ($\dot{r}_b$) of the scattering centers, is $f_{D_b}[n] = \frac{2\dot{r}_b[n]}{\lambda}$. We model the noise $z[n]$ as an additive circularly symmetric white Gaussian noise of variance $N_p$. From (\ref{eq:RxSig}), the received signal $x_{R}[n]$ is the sum of the convolution of the scattered signal from each $b^{th}$ point and the transmitted signal. The primitive-based technique, presented here, is computationally efficient and relatively accurate in generating m-D signatures and HRRPs. However the method does not capture the multipath effects. Algorithm~1 summarizes our simulation methodology.
%-------------------------------------------------------------------------------------------------
    	\begin{algorithm}[ht]
     \label{alg:SimMethod}
		\begin{algorithmic}[1]
        	\caption{Steps to generate 802.11ad-based radar signatures of an extended dynamic target}
        	\Statex \textbf{Input}: 
        	%\textit{Transmit Signal Model}: Transmitted signal for a PRI corresponding to $p^{th}$ packet  consisting of 512 fast time samples
            %\begin{equation}
            %    x(t) = \sqrt{E_s}\sum\limits_{n=0}^{511} G_{p,512}[n]h_T(t-nT_c)e^{j2\pi f_ct}. \nonumber
            %\end{equation} 
            %where
            %    \begin{equation}
            %        h_T[n]=h_T(nT_c)=\begin{dcases}
            %        1,\;n=0\\
            %        0,\;n\neq 0\\
            %        \end{dcases}\nonumber
            %    \end{equation} 
        	\textit{Dynamic extended target model}, with $B$ scattering centers during the $p^{th}$ PRI: animation model of car and bicycle from pyBullet, animation model of pedestrians from MoCap for $B$ point scatterers for $p=1$ to $P$ pulses $r_{b}[p]$, $b = 1$ to $B$ over the $m^{th}$ CPI.
            \Statex \textbf{Output:} Range-time profile $\chi^{RT}_m[r_n]$, range-Doppler ambiguity function $\chi^{RT}_m[r_n,f_D]$, and Doppler-time spectrogram $\chi^{DT}_m[f_D]$.
            \State Render each target into $B$ %multiple 
            elemental shapes or primitives:
            triangles for car, cylinders for bicycle, and ellipsoids for pedestrian. % is rendered into triangular facets; bicycle into cylindrical facets; and human into ellipsoids.
            \For{$p=0$ to $P-1$ packets of $m^{th}$ CPI} 
            \For{$b=1$ to $B$ and $n=0$ to $511$}
            \State \textit{(Target parameters)}: Compute aspect angle $\theta_b[n]$ of primitive of target with respect to radar
            \State Compute RCS $\sigma_b[n]$ for the shape of primitive.
            \State Compute reflectivity of a primitive\par\noindent\small
            \begin{equation*}
            a_b[n] =\zeta_b[n] \Gamma(\theta_b[n]) \sqrt{\sigma_b[n]} \frac{e^{-2\alpha r_b[n]}}{r_b^2[n]},
            \end{equation*}\normalsize
            Doppler-shift $f_{D_b}[n] = \frac{2\dot{r}_b[n]}{\lambda}$ and
            range-dependent time delay
            $\tau_b[n]=\frac{2r_b[n]}{cT_c}$.
            \EndFor
            \State \textit{(Noise)}: Generate 512 samples of additive circular-symmetric white Gaussian noise $z[n]\sim \mathcal{CN}(0,N_p)$.
			\State \textit{(Received signal)}: Compute the received signal from extended target over $m^{th}$ CPI comprising $P$ packets\par\noindent\small
            \begin{equation}
                x_R[n]= \sum\limits_{p=0}^{P-1}\sum\limits_{b=1}^{B} a_b[n] x_T(nT_c-\tau_b-pT_p)e^{-\mathrm{j}2\pi f_{D_{b}}pT_p} + z[n] \nonumber
            \end{equation}\normalsize
            \EndFor
            \State Compute high range resolution profile (HRRP) for $m^{th}$ CPI
            $\chi^{RT}_m[r_n]=\frac{1}{512P}\sum_{p=0}^{P-1}\left(x_R[n] \ast G_{p,N}[-n]\right)$ for $r_n = \frac{ncT_c}{2}, n=0:511.$
            \State Compute range-Doppler ambiguity function (AF)
            $\chi^{RT}_m[r_n,f_D]=\frac{1}{512P}\sum_{p=0}^{P-1}\left(x_R[n] \ast G_{p,N}[-n]\right)e^{\mathrm{j}2\pi f_D pT_p}$ for $f_D \in [-\nu_{max},\nu_{max}].$
            \State Compute Doppler-time spectrograms 
            $\chi^{DT}_m[f_D]=\sum_{n=0}^N\chi^{RT}_m[r_n,f_D]$.
		\end{algorithmic}
	\end{algorithm}

\section{Experiments}
\label{Sec:Experiments}
%--------------------------------------
\begin{figure}
\centering
\includegraphics[width=1.0\columnwidth]{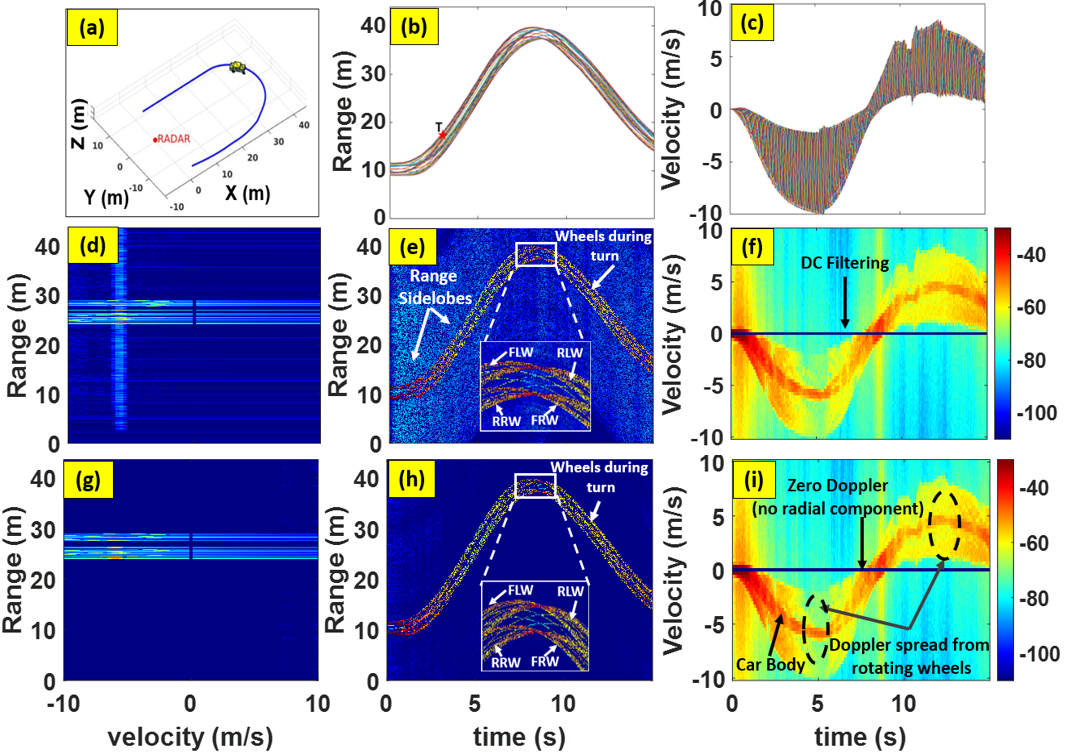}
\caption{(a) Ground truth trajectory of the car before the radar, (b) ground truth range time plots of point scatterers on the car and (c) ground truth radial velocity versus time of point scatterers on the car. (d)-(f) SG radar range-Doppler ambiguity plots at time instant T where the Doppler is maximum, range-time and velocity-time signatures, respectively. (g)-(i) MG range-Doppler ambiguity plots at time instant T, range-time and velocity-time signatures, respectively.}%\vspace{-12pt}}
\label{fig:CarSignaturesSim}
\end{figure}
%--------------------------------------------
We evaluated our approach %through numerical experiments 
for three common automotive targets - a small car, a bicycle and a pedestrian. We compared the results for both SG and MG waveforms. The noise variance in all experiments is $-100$ dBm.
\subsection{Car}
\label{subsec:car}
%\textbf{Car}: 
With the radar at the origin, we considered a trajectory of the car as shown in Fig.~\ref{fig:CarSignaturesSim}a. We model the car as a spatially large three-dimensional target and spans $4.2$ m $\times 2$ m $\times 2.2$ m. The automotive radar operates at $77$ GHz corresponding to a wavelength of $\lambda = 5.0$ mm.  The car accelerates from start, moves along a straight line to the left of the radar and then performs a U turn and moves towards the radar on a path to its right and then decelerates to a halt. The car is always within the maximum unambiguous range of the radar. Due to the complex trajectory undertaken by the car, the radar aspect varies considerably during the course of the motion (from $0^{\circ}$ to $180^{\circ}$). Fig.~\ref{fig:CarSignaturesSim}b shows the ground truth of the range-time for different point scatterers situated on the moving car. The range increases as the car moves away from the radar and then make a U turn at $t=8$ s after which the range again begins to reduce. Fig.~\ref{fig:CarSignaturesSim}c shows the ground truth of the velocity time behaviour of the point scatterers. We notice that there is considerable variation in the velocities of the different point scatterers on the body of the car and especially from the wheel. The Dopplers of different points on the chassis of the car also show variation depending on their proximity and aspect with respect to the radar.

We present the range-Doppler ambiguity plot, the high range resolution profile and the Doppler spectrograms for both the SG and MG waveforms. A notch filter has been implemented at zero Doppler similar to the measurement data. The range-Doppler AF plots of the car, (shown in Figs.~\ref{fig:CarSignaturesSim}d and g) are computed when the car's Doppler is maximum at time instant $5$ s. The plots show multiple point scatterers corresponding to the different parts of the car. The waveform is characterized by a Doppler ambiguity of 0.3 m/s and a range ambiguity of 0.085 m. The Doppler velocity is shown only for the span [-10 m/s, +10 m/s] and we observe high Doppler sidelobes here for both SG and MG which arise from Fourier processing. % arising from the Fourier processing. 
The range axis is plotted up to the maximum unambiguous range. In Fig.~\ref{fig:CarSignaturesSim}d, the SG waveform shows significantly high sidelobes in the range dimension. This is due to high Doppler velocities of multiple point scatterers that perturb the perfect autocorrelation property of Golay complementary pair. In contrast, as illustrated in Fig.~\ref{fig:CarSignaturesSim}g, the MG waveform is free of sidelobes in the range dimension. This demonstrates that waveform is Doppler-resilient and retains perfect autocorrelation along the range dimension. 

The range-time plots for SG and MG in Fig.~\ref{fig:CarSignaturesSim}e and h, show very good agreement with the ground truth range-time plot. In the inset, we observe the m-R features that arise from the micro-motions of the different point scatterers on the car. Due to the effect of shadowing, the range tracks show some discontinuities. The range resolution of the radar is sufficient at some time instances in resolving these m-R tracks. Similar to the range-Doppler AF, the SG signature shows significant range sidelobes due to the motion of the car. These sidelobes are absent in the case of the MG signatures. The velocity-time results (Fig.~\ref{fig:CarSignaturesSim}f and i) for both standard and modified Golay agree with the ground truth velocity-time plots. The m-D tracks of the different point scatterers can be easily observed. Results for SG and MG are nearly identical in this case. This is because the motion of the car only affects the range dimension and not the Doppler dimension. 

\begin{figure}[t]
     \centering
     \includegraphics[width=0.60\columnwidth]{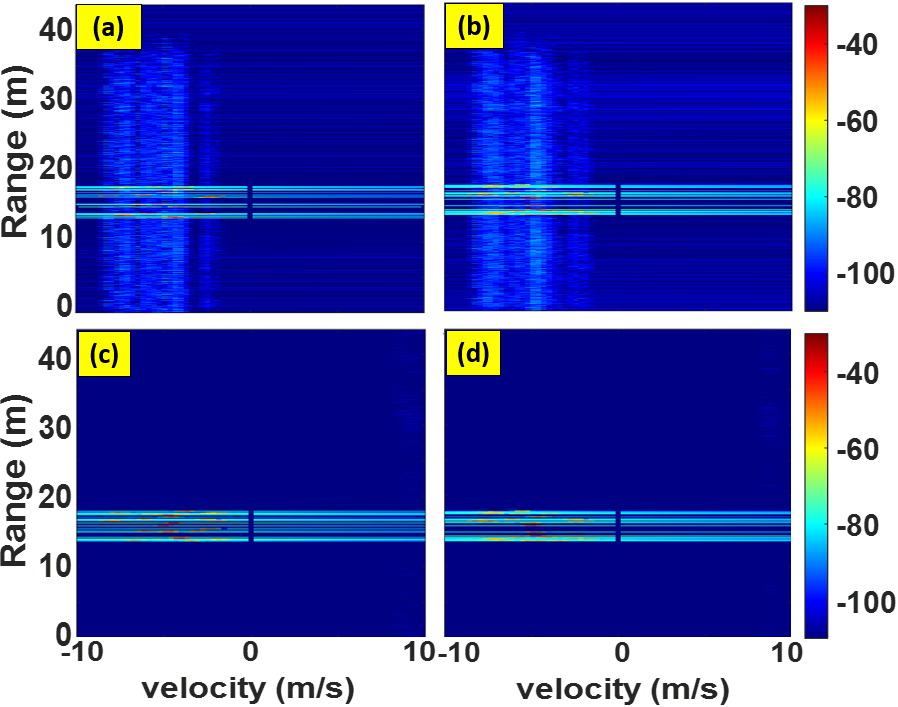}
     \caption{SG radar range-Doppler ambiguity plots with (a) randomly selected scatterers, and (b) with shadowing taken into account, respectively.  MG range-Doppler ambiguity plots with (c) randomly selected scatterers, and (d) with shadowing taken into account, respectively.} %\vspace{-12pt}}
     \label{fig:car_scattering}
     \end{figure}

%\textcolor{blue}{If the software is ``in-house'', then what is the reference [42] for?}
To indicate a primitive's visibility, we performed the ray tracing for the car using a custom-developed ray tracing software based on shooting and bouncing technique \cite{ling1989shooting}. However, this is repeated every PRI leading to high computational cost. It is, therefore, usually employed for static scenarios. On the other hand, the stochastic method based on the random Bernoulli-distributed variable is computationally more efficient. We generated a single CPI range-Doppler AF plot for both SG (Fig.~\ref{fig:car_scattering}b) and MG (Fig.~\ref{fig:car_scattering}d) sequences. We compared these with the corresponding range-Doppler AFs generated using the random Bernoulli-distributed variable for waveforms in Figs.~\ref{fig:car_scattering}a and c, respectively.
%To demonstrate that the stochastic model of determining a primitive's visibility results in images close to reality, we derive ground truth results using an in-house custom-developed software based on shooting and bouncing technique \cite{ling1989shooting}. The ray tracing method is extensively used for high frequency radar modeling of large targets. However, the method is computationally intensive in terms of both time and memory and hence is largely used in static scenarios. The stochastic method, based on the random Bernoulli-distributed variable, on the other hand, is computationally quick and simple. We generate the range-Doppler AF plot for one coherent processing interval for both SG and MG sequences using both methods and show them in Fig.\ref{fig:car_scattering}.} 
	 %----------------------------------------
 %----------------------------------------

Qualitatively, stochastic illumination yields signatures (Figs.~\ref{fig:car_scattering}a and c) that are closer to the images generated by ray tracing (Figs.~\ref{fig:car_scattering}b and d). We quantified the performance by computing normalised-mean-squared-error (NMSE) and structural similarity index metric (SSIM) \cite{hore2010image}. When two images are similar, their NMSE vanishes and SSIM approaches unity. The NMSE of range-Doppler AFs for SG and MG waveforms are $0.18\%$ and $0.17\%$, respectively, while both SSIMs are unity. This implies that the AFs are almost identical.%\\
\subsection{Bicycle}
\begin{figure}
\centering
\includegraphics[width=1.0\columnwidth]{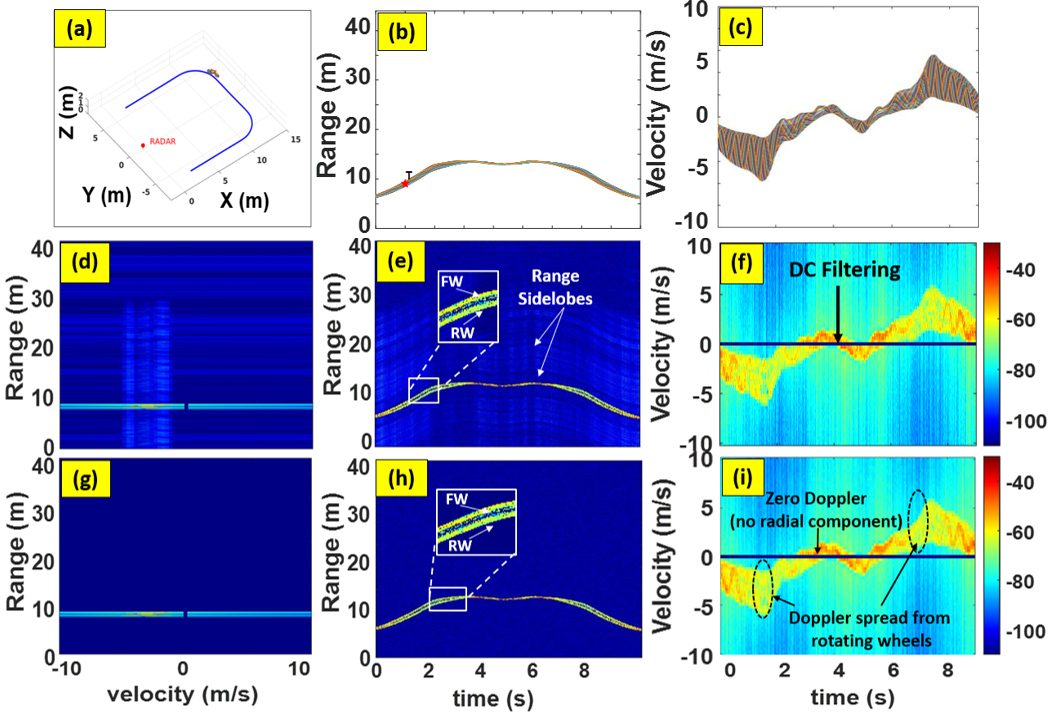}
\caption{(a) Ground truth trajectory of the bicycle before the radar, (b) ground truth range time plots of point scatterers on the bicycle and (c) ground truth radial velocity versus time of point scatterers on the bicycle. (d)-(f) SG radar range-Doppler ambiguity plots, range-time and velocity-time signatures, respectively. (g)-(i) MG range-Doppler ambiguity plots, range-time and velocity-time signatures, respectively.} %\vspace{-12pt}}
\label{fig:BicycleSignaturesSim}
\end{figure}

%\textbf{Bicycle}:
We consider a bicycle moving along the trajectory shown in Fig.~\ref{fig:BicycleSignaturesSim}a. 
%--------------------------------------------
%--------------------------------------------
The bicycle accelerates from halt and reaches a steady velocity and then performs two right turns before halting. The m-R and m-D features from the motion of the bicycle are presented in Fig.\ref{fig:BicycleSignaturesSim}d-i. The ground truth range-time plots of the different point scatterers on the two wheels show a very narrow range spread except during the turns (at 1.5 s and 7 s) - especially in comparison to the car. We compute the range-Doppler AFs at 1 s when the Doppler velocity is maximum. The AF plot for SG in Fig.~\ref{fig:BicycleSignaturesSim}d shows high sidelobes along the range dimension due to the Doppler-shifted point scatterers. The AF plot for the MG in Fig.~\ref{fig:BicycleSignaturesSim}g shows no sidelobes along the range because of the resilience of the waveform to Doppler shifts. The HRRPs for SG in Fig.~\ref{fig:BicycleSignaturesSim}e and MG waveforms in Fig.~\ref{fig:BicycleSignaturesSim}h are very similar to the ground truth plots. However, the high range sidelobes are evident in the SG plots. The velocity-time plots in Fig.~\ref{fig:BicycleSignaturesSim}f and i, show significantly greater Doppler spread due to the rotating wheels. The spread is greatest during turns which is similar to the results from the experimental measurements. The results from SG and MG look nearly identical here and are very similar to the ground truth results. 
\subsection{Pedestrian}
\label{subsec:ped}
\begin{figure}
\centering
\includegraphics[width=0.9\columnwidth]{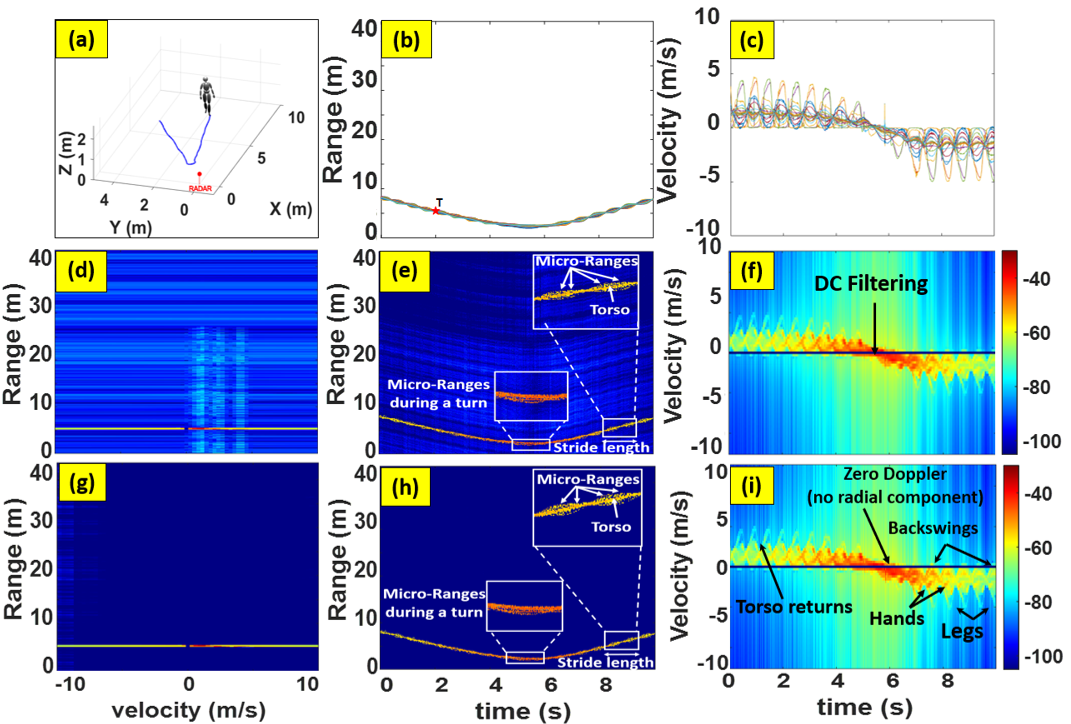}
\caption{(a) Ground truth trajectory of the human before the radar, (b) ground truth range time plots of point scatterers on the human and (c) ground truth radial velocity versus time of point scatterers on the human. (d)-(f) SG radar range-Doppler ambiguity plots, range-time and velocity-time signatures, respectively. (g)-(i) MG range-Doppler ambiguity plots, range-time and velocity-time signatures, respectively.} %\vspace{-12pt}}
\label{fig:HumanSignatures}
\end{figure}

%\textbf{Pedestrian}: 
Next, we study the radar signatures of the pedestrian in Fig.~\ref{fig:HumanSignatures}. 
%--------------------------------------------
%--------------------------------------------
The trajectory followed by the subject is shown in Fig.~\ref{fig:HumanSignatures}a. Here the pedestrian approaches the radar and then turns around and walks away from the radar. The range-Doppler AFs are shown for time instant of 2s when the micro-Dopplers peak. The SG range-Doppler AF (Fig.~\ref{fig:HumanSignatures}d) shows sidelobes along the range which are absent in the AF for MG (Fig.~\ref{fig:HumanSignatures}g). We observe again that the range-time plots for SG (Fig.~\ref{fig:HumanSignatures}e) and MG (Fig.~\ref{fig:HumanSignatures}h) sequences are in agreement with the ground truth results (Fig.~\ref{fig:HumanSignatures}b). Due to the smaller spatial extent of the pedestrian across the range dimension, the m-R tracks are difficult to observe except at some time instants. The inset shows the m-R from the right and left legs and arms. The pedestrian is always within the maximum unambiguous range and the field of view of the radar.  The Doppler velocity-time spectrograms in Fig.~\ref{fig:HumanSignatures}f and i show excellent agreement with the ground truth results Fig.~\ref{fig:HumanSignatures}c. We observe the m-Ds from the feet, legs, arms and torso. The Dopplers are positive when the pedestrian approaches the radar and are negative when the pedestrian moves away from the radar. The strongest Dopplers arise from the torso. As observed before for car, the signatures from the SG and MG are nearly identical in the Doppler domain.
\subsection{Multiple Targets}
\label{subsec:Multi_tar}
\begin{figure}
    \centering
    \includegraphics[width=1.0\columnwidth]{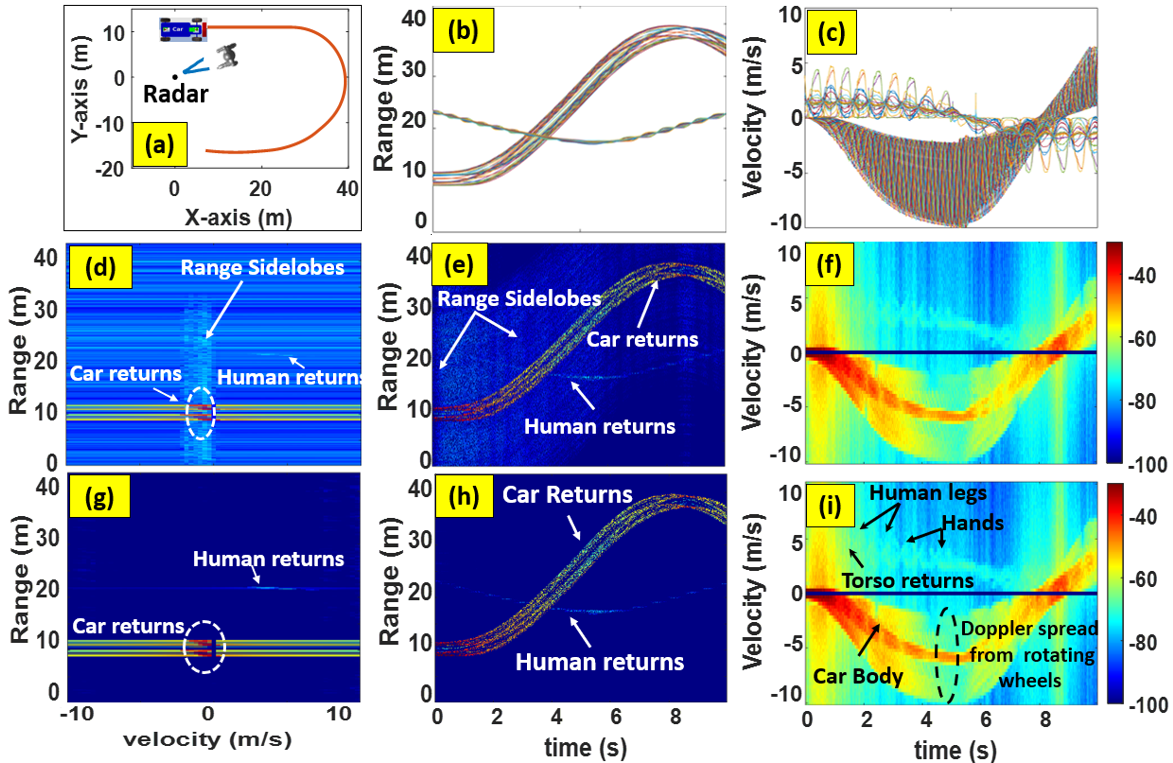}
    \caption{(a) Overlaid ground truth trajectories of a pedestrian and a car, (b) ground truth range-time plots and (c) ground truth radial velocity versus time for the signal reflected off from the point scatterers on the human and a car. (d)-(f) SG radar range-Doppler ambiguity plots, range-time and velocity-time signatures, respectively. (g)-(i) MG range-Doppler ambiguity plots, range-time and velocity-time signatures, respectively.} %\vspace{-12pt}}
    \label{fig:MultiTarSignatures}
    \end{figure}

%\textbf{Multiple Targets}:
Next, we consider a scenario with multiple targets. Fig.~\ref{fig:MultiTarSignatures}a shows the trajectories followed by a car and a pedestrian within the common radar coverage area. The car moves away from the radar and then turns around and approaches the radar while the pedestrian moves towards the radar from a $20$ m range and then turns and walks away. The received radar signal is the superposition of the scattered signals from the two targets along with noise. The RCS of the pedestrian is lower than that of the car. We computed the range-Doppler AFs at $1$ s when the car is closest to the radar. The range-Doppler AF for SG and MG waveforms are shown in Fig.~\ref{fig:MultiTarSignatures}d and Fig.~\ref{fig:MultiTarSignatures}g, respectively. 

This clearly shows that, with the strong range sidelobes of the car, the relatively weaker pedestrian target is difficult to detect in Fig.~\ref{fig:MultiTarSignatures}d with the SG waveform. However, the improvement in the peak-to-sidelobe level in the MG waveform results in a clearly visible human in Fig.~\ref{fig:MultiTarSignatures}g. We next observed the range-time signatures for SG (Fig.~\ref{fig:MultiTarSignatures}e) and MG (Fig.~\ref{fig:MultiTarSignatures}h) sequences. We notice that the range trajectories are in perfect agreement with the ground truth results of Fig.~\ref{fig:MultiTarSignatures}b. Again, in the case of SG waveform, a pedestrian is difficult to discern in the presence of high range sidelobes in Fig.~\ref{fig:MultiTarSignatures}e. This is not so with the MG waveform, wherein the use of a Doppler-resilient sequence suppresses the range sidelobes. The Doppler velocity-time spectrograms for SG (Fig.~\ref{fig:BicycleSignaturesSim}f) and MG (Fig.~\ref{fig:MultiTarSignatures}i) waveforms show radar backscatter from both the targets with corresponding micro-Doppler features. The spectrograms are very similar to the ground truth plots in Fig.~\ref{fig:MultiTarSignatures}c. % In the next section, we discuss the quantitative detection results of the two targets for both the SG and MG waveforms.
\section{Detection performance}
\label{sec:Detection}
%\textbf{Detection performance}:
%-------------------------------------------------------
%------------------------------------------------------------------------------------    
%------------------------------------------------------
% \begin{figure*}%[t]
% \centering 
% \includegraphics[width=1\textwidth]{combined_detection_strategy_new5.PNG}
% % \includegraphics[width=0.9\columnwidth]{combined_detection_strategy_new4.pdf}
% \caption{Histogram of car, pedestrian and noise signals for SG and MG (a) at SNR of +5dB and (b) at SNR of -20dB; (c) The $P_{\textrm{d}}$ for SG and MG for different SNR values at fixed $P_{\textrm{fa}}=10^{-6}$. (d) The $P_{\textrm{fa}}$ for SG and MG as a function of SNR of radar receiver.}
% \label{fig:RadarPerformance}
% \end{figure*}
\begin{figure}%[t]
\centering 
\includegraphics[width=1\columnwidth]{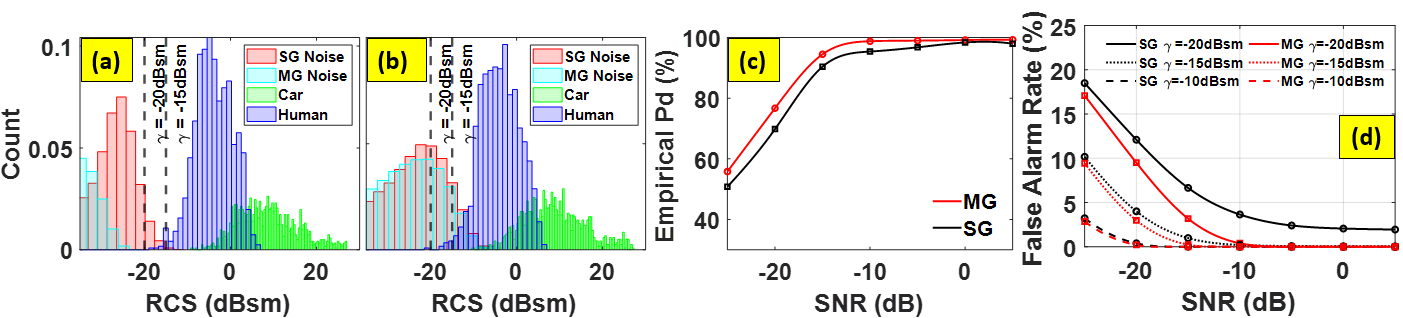}
\caption{Histogram of car, pedestrian and noise signals for SG and MG (a) at SNR of +5dB and (b) at SNR of -20dB; (c) The $P_{\textrm{d}}$ for SG and MG for different SNR values at fixed $P_{\textrm{fa}}=10^{-6}$. (d) The $P_{\textrm{fa}}$ for SG and MG as a function of SNR of radar receiver.}
\label{fig:RadarPerformance}
\end{figure}
%-------------------------------------------------------------------------------------------------
Let $\bm{\psi} = (a_b, \tau_b, f_{D_b}) \in \Psi = \mathbb{C} \times [0, T_p] \times [-2\nu_{\textrm{max}}/\lambda, 2\nu_{\textrm{max}}/\lambda]$ be the unknown parameters of the received signal \eqref{eq:RxModel} in the parameter space $\Psi$, where we have omitted the time-dependency of RCS for simplicity. We define $\mathcal{H}_1$ as the composite hypothesis for the presence of target and $\mathcal{H}_0$ as the null hypothesis,\par\noindent\small
\begin{align}
\label{eq:hyp}
\mathcal{H}_0:\;x_R[n]  &=z[n],\;n=0,\cdots,N,\nonumber\\
\mathcal{H}_1:\;x_R[n]  &= \sum\limits_{p=0}^{P-1}\sum\limits_{b=1}^{B} a_b s_T(nT_c-\tau_b-pT_p)e^{-\mathrm{j}2\pi f_{D_{b}}pT_p} + z[n],\nonumber\\
&\;\;\;\;\;\;\;\;\;\;\;\;\;\;n=0,\cdots,N.
\end{align}\normalsize
Given fixed ${\bm{\psi}} \in \Psi$, we define $\Gamma({\bm{\psi}})$ as the log-likelihood ratio (LLR) between $\mathcal{H}_1$ and $\mathcal{H}_0$. The binary hypothesis testing \eqref{eq:hyp} is solved via generalized likelihood ratio test (GLRT),\par\noindent\small
\begin{align}
    \underset{{\bm{\psi}} \in \Psi}{\textrm{max}} \Gamma({\bm{\psi}}) \LRT{H_0}{H_1} \gamma,
\end{align}\normalsize
where $\gamma$ is the detection threshold. Since $z[n]\sim \mathcal{CN}(0,N_p)$, the probability distributions in both hypothesis are known. The LLR is
\par\noindent\small
\begin{align}
    \Gamma({\bm{\psi}}) = \ln L_N^{(\mathcal{H}_1)}/L_N^{(\mathcal{H}_0)},
\end{align}\normalsize
where $L_N^{(\mathcal{H}_i)} = \prod_{n=1}^Nf( x_R,\mathcal{H}_i) $ is the likelihood function of the samples $x_R[0], \cdots, x_R[N-1]$ under the hypothesis $\mathcal{H}_i$ and $f(\cdot, \cdot)$ is the probability density function of the complex Gaussian distribution. Expanding the LLR for each $b^{th}$ scatterer yields \cite{poor1994introduction}\par\noindent\small 
\begin{align}
    \Gamma({\bm{\psi}}) = 2\operatorname{\mathbb{R}e}\left( \frac{a_b^{\ast}}{N_p}\sum\limits_{p=0}^{P-1}x_R[n]*G_{p,N}[-n]\right) - \frac{a_b^2P_t}{N_p},
\end{align}\normalsize
where the function $\operatorname{\mathbb{R}e}(\cdot)$ provides the real part of the argument and $(\cdot)^\ast$ denotes the complex conjugate. Fixing the $\tau_b$ and $f_{D_b}$, $\Gamma({\bm{\psi}})$ is maximized to obtain $a_b$ as\par\noindent\small
\begin{align}
a_b = \frac{1}{P_t} \sum\limits_{p=0}^{P-1}x_R[n]*G_{p,N}[-n].
\end{align}\normalsize
Substituting this $a_b$ produces the following GLRT \par\noindent\small
\begin{align}
\label{eq:cohdet}
\underset{\substack{\tau_b\in [0, T_p],\\f_{D_b}\in [-2\nu_{\textrm{max}}/\lambda, 2\nu_{\textrm{max}}\lambda]}}{\textrm{max}} \dfrac{1}{N_pP_t}\left|  \sum\limits_{p=0}^{P-1}x_R[n]*G_{p,N}[-n] \right|^2 \LRT{H_0}{H_1} \gamma.
\end{align}\normalsize
From this relation, it follows that the detection performance of MG signals is superior over a wider Doppler range $[-\nu_{\textrm{max}}, \nu_{\textrm{max}}]$ because $\chi^{RD}_m[r_n,f_D]$ is based on the correlation of Golay pairs (cf. (\ref{eq:RDAmbig})).

We validated this performance experimentally. From the minimum possible RCS ($-30$ dBsm) and the maximum range ($43$ m) of the radar, we estimate the minimum detectable signal of the radar to be $-100$ dBm from (\ref{eq:RxSig}). We define the minimum SNR of the radar receiver as the ratio between the minimum detectable signal of the radar and the average noise power ($N_p$). In our study, we vary this SNR from $-25$ to $+5$ dB. We consider a scenario where both car and pedestrian move simultaneously before the radar following the trajectories shown in the previous section. The received radar signal is therefore the superposition of the scattered signals from the two targets along with noise. In order to study the detection performance of the radar, we consider the radar range-time results ($\chi^{RT}_m[r_n]$) where $m$ denotes the CPI and $r_n$ denotes the discrete range bin. We multiply the signal at every bin with the quadratic power of the corresponding range. 
This step is crucial while detecting multiple targets of differing cross-sections because it removes the dependency of signal strength on the distance of the target from the radar.

The extended targets are spread across multiple range bins and the RCS of a target (car or pedestrian) at each CPI is obtained by the \emph{coherent integration} of the range compensated signal across multiple range bins determined by the ground truth range-time plots (the rest of the range bins have noise). Note that the statistic $\sum\limits_{p=0}^{P-1}x_R[n]*G_{p,N}[-n]$ in \eqref{eq:cohdet} is the output of the coherent signal integrator in the range-Doppler-time dimension cube. So, we further integrate this output along the Doppler dimension to obtain $\chi^{RT}_m[r_b]$ and compensate $a_b$ for range using \eqref{eq:RCSellipsoid}. The RCS of, for instance, the car $\sigma_{\text{car}}$ at every $m^{th}$ CPI is then\par\noindent\small
\begin{equation}
\label{eq:RangeCompensation}
    \sigma_{\text{car}}(m)=\left\lVert\sum\limits_{b=1}^B \chi^{RT}_m[r_b]r_b^2e^{\mathrm{j}2\frac{2\pi}{\lambda}r_b}\right\rVert^2,
\end{equation}\normalsize
where $\{r_b\}_{b=1}^B$ denotes range bins determined from the ground truth car data for that CPI. In Fig.~\ref{fig:RadarPerformance}, we plot the histogram distribution of the noise and target returns for both SG and MG sequences, from all the CPIs, under two different SNRs: $-20$ and $+5$ dB. The histograms of the car and pedestrian RCS do not show significant variation for SG and MG. Hence, we show a single distribution for each of these targets. It is evident that the RCS of the car fluctuates from $-10$ dBsm to $30$ dBsm with a mean of $10$ dBsm. In case of pedestrian, the RCS is in the range $-20$ to $+5$ dBsm. The variation in RCS arises from the change in aspect angle with respect to the radar. As expected, a car thus has noticeably higher RCS than a pedestrian. Under poor SNR conditions, noisy returns escalate. Both histograms show that such returns are higher for SG than MG because, in case of the former, noise is added to the high range sidelobes. 

The empirical $P_\textrm{d}$ of the radar is estimated from the area under the target histograms beyond the RCS threshold, $\gamma$ (indicated by the dashed line in the Fig.~\ref{fig:RadarPerformance}). Similarly, the empirical $P_{\textrm{fa}}$ is determined from the area under the noise histograms above the same threshold. 
We examined the $P_{\textrm{fa}}$ in Fig.~\ref{fig:RadarPerformance}d for three different thresholds. With the increase in threshold, $P_{\textrm{fa}}$ decreases. When $\gamma=-15$ dBsm, SG exhibits higher $P_{\textrm{fa}}$ than MG by approximately 2.5\% for low SNRs ($-20$ to $0$ dB). This is even more pronounced for the $-20$ dBsm threshold curve because the high SG range sidelobes with additive noise show up as false alarms. 
%When the noise is very high (SNR is $-30$ dB), the noise peaks are above the range sidelobe levels. Hence the $P_{\textrm{fa}}$ of SG and MG become similar. 
For moderate-to-high SNR, i.e. $0$ to $+5$ dBsm, $P_{\textrm{fa}}$ for SG is higher than MG for $-20$ dBsm threshold. 
%------------------------------------------------------
% \begin{figure*}[t]
% \centering 
% \includegraphics[scale=1]{RadarROC.png}
% \caption{(Histogram of car, pedestrian and noise signals for standard and modified Golay (a) at SNR of +5dB and (c) at SNR of -20dB. The Pd for standard Golay (SG) and modified Golay (MG) for different threshold values. The false alarm rate for SG and MG as a function of signal to noise ratio of radar receiver.}
% \label{fig:RadarPerformance}
% \end{figure*}
% \begin{figure*}[t]
% \centering 
% \includegraphics[scale=0.5]{combined_detection_strategy.pdf}
% \caption{Histogram of car, pedestrian and noise signals for standard and modified Golay (a) at SNR of +5dB and (c) at SNR of -20dB. The Pd for standard Golay (SG) and modified Golay (MG) for different threshold values. The false alarm rate for SG and MG as a function of signal to noise ratio of radar receiver.Better to put all four subfigures in a single row. It will save space. Currently, the space around the sides is being wasted.} }
% \label{fig:RadarPerformance}
% \end{figure*}
%-----------------------------------------------------------------
\begin{table}[t]
\centering
\caption{Minimum SNR required for achieving desired $P_\textrm{\normalfont d}$ at $P_{\textrm{\normalfont fa}}=10^{-6}$}
\label{Table:SNR_com}
\begin{tabular}{|c|c|c|c|c|c|}
\hline
$P_d$ ($\%$) & $95$ & $96$ & $97$ & $98$ & $99$ \\ \hline
SG SNR (dB) & $-11.5$ & $-7.2$ & $-4.1$ & $-1.35$ & $1.9$ \\ \hline
MG SNR (dB) & $-14.75$ & $-14$ & $-13.4$ & $-12.5$ & $-9.4$ \\ \hline
\end{tabular}
\end{table}
%-----------------------------------------------------------------

In a standard radar operation, thresholds are usually set to achieve a desired probability of detection for a fixed value of false alarm rate. We choose a low $P_{\textrm{fa}} = 10^{-6}$ and obtained corresponding $P_\textrm{d}$ for different SNRs in Fig.~\ref{fig:RadarPerformance}c. This result clearly indicates a superior detection performance of MG over SG, especially at low SNR regime. Further, Table~\ref{Table:SNR_com} lists the minimum SNRs that the radar can tolerate to achieve a specified $P_\textrm{d}$. We notice that the inherent Doppler resilience of MG waveform leads to achieving the same $P_\textrm{d}$ at much lower SNR levels than the SG waveform. For example, a $P_\textrm{d}=99\%$ is maintained by the radar at SNR of $-9.4$ and $1.9$ dB while transmitting MG and SG, respectively. Thus, the performance improvement margin with MG is \texttildelow$11$ dB in SNR over SG.
\section{Summary}
\label{sec:Conclusion}
We presented a USRR in the context of recent advances in joint radar-communications system. which employs the 512-bit Golay codes in 802.11ad link for range estimation up to $40$ m with a resolution of $0.085$ m. The codes in consecutive packets form Golay complementary sequences based on the PTM sequence that results in very low sidelobe levels for most automotive targets moving up to $144$ km/hr. We demonstrated detection of HRRPs and m-D spectrograms of common automotive targets - pedestrian, bicycle and car. Each of these targets were animated and modeled as extended targets with multiple scattering centers distributed along their body. The signatures from the targets show distinctive micro-motion features such as the rotation of the wheels and the swinging motions of the arms and legs. The detection performance of the radar shows a marked reduction in the $P_{\textrm{fa}}$ for the MG when compared to SG for low and moderate SNRs. The MG waveform maintains the same detection performance as SG even at much lower SNR values.

\section*{Acknowledgments}
G. D., S. V. and S. S. R. were funded by the Air Force Office of Scientific Research (AFOSR), Asian Office of Aerospace Research and Development (AOARD) through the 5IOA036 FA23861610004 grant. %K. V. M. was supported by Iowa Flood Center. 
K. V. M. acknowledges discussion with Prof. Robert A. Calderbank of Duke University regarding the latter's work in \cite{chi2011complementary}.

%\balance
\bibliographystyle{main}
\bibliography{refs}

\end{document}